\documentclass[twocolumn]{aastex6}
\usepackage{natbib}
\usepackage{color}

\newcommand     \mum    {\,\mu{\rm m}}  

\def \bea {\begin{eqnarray}}
\def \ena {\end{eqnarray}}

\def    \abs     {\rm  abs}

\def	\B	{{\rm B}}

\def	\bJ	{{\bf J}}
\def	\bk	{{\bf k}}

\def    \bmu    {{\hbox{\boldsym\char'026}}}	

\def    \bomega {{\hbox{\boldsym\char'041}}}	

\def	\cm	{\,{\rm cm}}
\def	\d	{{\rm d}}

\def	\eff	{{\rm eff}}
\def	\ed	{{\rm ed}}
\def	\ehat	{\hat{\bf e}}
\def	\erg	{\,{\rm ergs}}

\def	\gas	{\,{\rm gas}}

\def	\H	{{\rm H}}

\def	\LTE	{{\rm LTE}}

\def	\IR	{{\rm IR}}

\def	\Jy	{\,{\rm Jy}}
\def	\K	{{\rm K}}
\def    \kB    {k_{\rm B}}

\def	\s	{\,{\rm s}}

\def	\sr	{\,{\rm sr}}
\def	\rot	{{\rm rot}}

\def	\tot 	{\rm {tot}}
\def	\rad 	{\rm {rad}}

\def	\Bar	{{\rm Bar}}

\def	\xhat		{\hat{\bf x}}
\def	\yhat		{\hat{\bf y}}
\def	\zhat		{\hat{\bf z}}
\def	\ahat		{\hat{\bf a}}
\def	\ehat		{\hat{\bf e}}

\def    \Bv     	{\bf  B}

\def    \kv     	{\bf  k}

\def	\erg			{{\rm erg}}

\font\mib=cmmib10

\def\bmu{\hbox{\mib\char"16}}

\def\bomega{\hbox{\mib\char"21}}

\begin{document}
\title{Cross section alignment of polycyclic aromatic hydrocarbons by anisotropic radiation}
\author{Thiem Hoang}
\affil{Korea Astronomy and Space Science Institute, Daejeon 34055, Korea, email: thiemhoang@kasi.re.kr}
\affil{Korea University of Science and Technology, Daejeon, 34113, Korea}
\author{A. Lazarian}
\affil{Astronomy Department, University of Wisconsin, Madison, WI 53706}

\begin{abstract}
We study the effect of anisotropic radiation illumination on the alignment of polycyclic aromatic hydrocarbons (PAHs) and report that cross-sectional mechanism of alignment earlier considered in terms of gas-grain interactions can also be efficient for the photon-grain interaction. We demonstrate this by first calculating the angle-dependence rotational damping and excitation coefficients by photon absorption followed by infrared emission. We then calculate the degree of PAH alignment for the different environments and physical parameters, including the illumination direction, ionization fraction, and magnetic field strength. For the reflection nebula (RN) conditions with unidirectional radiation field, we find that the degree of alignment tends to increase with increasing the angle $\psi$ between the illumination direction and the magnetic field, as a result of the decrease of the cross-section of photon absorption with $\psi$. We calculate the polarization of spinning PAH emission using the obtained degree of alignment for the different physical parameters, assuming constant grain temperatures. We find that the polarization of spinning PAH emission from RN can be large, between $5-20~\%$ at frequencies $\nu > 20$ GHz, whereas the polarization is less than $3~\%$ for photodissociation regions (PDRs). In realistic conditions, the polarization is expected to be lower due to grain temperature fluctuations and magnetic field geometry. The polarization for the diffuse cold neutral medium (CNM) is rather low, below $1~\%$ at $\nu>20$ GHz, consistent with observations by WMAP and Planck. Our results demonstrate that the RNe are the favored environment to observe the polarization of spinning dust emission as well as polarized mid-IR emission from PAHs.
 \end{abstract}

\keywords{ISM: dust, extinction ---
          ISM: general ---
	  galaxies: ISM ---
          infrared: galaxies
	  }
\section{Introduction}
Polycyclic aromatic hydrocarbons (PAHs) is an important dust component of the interstellar medium (ISM, see \citealt{2008ARA&A..46..289T} for a review). PAH molecules are planar structures, consisting of carbon hexagonal rings and hydrogen atoms attached to their edge via valence bonds. PAHs are the leading carrier for the 2175\AA~broad feature in the extinction curve (\citealt{1989IAUS..135..313D}). Moreover, upon absorbing ultraviolet (UV) photons, PAHs reemit radiation in mid-infrared features, including 3.3, 6.2, 7.7, 8.6, 11.3, and 17 $\mu$m, due to vibrational transitions (\citealt{1984A&A...137L...5L}; \citealt{1985ApJ...290L..25A}; \citealt{2007ApJ...657..810D}). The evidence of PAHs is ubiquitously observed from our Galaxy (see, e.g., \citealt{2008ARA&A..46..289T}) and external galaxies (\citealt{2007ApJ...656..770S}) to circumstellar disks around young stars (\citealt{2017ApJ...835..291S}).

Rapidly spinning PAHs emit microwave radiation via a new mechanism, so-called spinning dust (\citealt{1998ApJ...508..157D}; \citealt{Hoang:2010jy}). The latter is the most likely origin of anomalous microwave emission (AME) that contaminates Cosmic Microwave Background (CMB) radiation  (\citealt{Kogut:1996p5293}; \citealt{Leitch:1997p7359}).\footnote{PAH molecules definitely emit microwave radiation. The efficiency of their emission depend on the dipole moments of the molecules and the latter parameter is somewhat uncertain. Therefore we do not know exactly what percentage of microwave radiation has the PAH origin. Alternative sources of AME include magneto-dipole emission from strongly magnetized classical large grains (\citealt{1999ApJ...512..740D}; \citealt{2013ApJ...765..159D}), magnetic nanoparticles (\citealt{2016ApJ...821...91H}), as well as silicate spinning nanoparticles (\citealt{2016ApJ...824...18H}; \citealt{2017ApJ...836..179H}).} Spinning dust emission can also be a source of excess microwave emission in some circumstellar disks around Herbig Ae/Be stars (\citealt{Hoang:2018td}).

The alignment of PAH molecules can leave the imprints in: (1) the UV polarization of starlight at 2175~\AA, (2) polarized mid-IR emission, and (3) polarized spinning dust emission. To date, only two stars exhibit the UV polarization bump, and inverse modeling in \cite{2013ApJ...779..152H} found that a considerable degree of PAH alignment with the magnetic field is required to reproduce the observed polarization bump. Observational studies of polarized PAH emission are still limited. Early search by \cite{1988A&A...196..252S} reports the detection of polarized PAH emission from the Orion ionization front, with $p\sim 0.86\pm 0.28 \%$ for the 3.3 $\mu$m feature, although it is cautious because the ground-based measurement at $3.3~\mum$ is very challenging due to the atmospheric opacity. Moreover, observational studies of the AME polarization (\citealt{2006ApJ...645L.141B}; \citealt{2009ApJ...697.1187M}; \citealt{2011MNRAS.418L..35D}; \citealt{Battistelli:2015dt}) only found the upper limits, with the polarization level between $1-5 \%$. An upper limit of $1 \%$ is also reported in various studies (see \citealt{LopezCaraballo:2011p508}; \citealt{RubinoMartin:2012ji}; \citealt{2017MNRAS.464.4107G}). Therefore, a physical prediction for polarized spinning dust emission is crucially important for interpretation of the observed AME polarization. 

Very recently, \cite{Zhang:2017ea} detected the polarization at 11.3$~\mu$m from the MWC 1080 nebula. The measured polarization of $1.9 \pm 0.2 \%$ can be successfully explained when PAHs can be aligned with the magnetic field to a degree of ~ $10~\%$. Yet an underlying mechanism of PAH alignment is still unclear.

The theory of grain alignment (see \citealt{LAH15} for a review) was developed to mainly explain the alignment of large classical grains (i.e., above $\sim 0.01~\mum$) rather than nanoparticles. For instance, the analytical theory of Radiative Torques (RATs) acting on ordinary paramagnetic grains (\citealt{2007MNRAS.378..910L}; \citealt{Hoang:2008gb}) as well as the theory of RAT acting on grains with magnetic inclusions (\citealt{Lazarian:2008fw}; \citealt{2016ApJ...831..159H}) do not predict any alignment of nanoparticles as the efficiency of RATs decreases rapidly as the size of the grain gets much smaller than the radiation wavelength. The mechanical alignment arising from the gas-grain interactions can occur, but for both the alignment of non-helical grains (see \citealt{1997ApJ...483..296L}) and the mechanical alignment of helical grains (\citealt{2007ApJ...669L..77L}; \citealt{2016MNRAS.457.1958D}; \citealt{2018ApJ...852..129H}), relatively high drift velocities of gas and dust are required. The mechanisms of dust acceleration involving Magneto-hydrodynamics (MHD) turbulence (see \citealt{Hoang:2012cx} and ref. therein) as well as due to Coulomb interaction of fluctuating charges on grains (\citealt{2012ApJ...761...96H}) do not provide sufficiently high drift velocities.  

\cite{2000ApJ...536L..15L} first suggest that rapidly spinning nanoparticles can be weakly aligned by {\it resonance paramagnetic relaxation}, a modified version of Davis-Greenstein mechanism \citep{1951ApJ...114..206D} that works in rapidly spinning tiny grains. Numerical calculations in \cite{2014ApJ...790....6H} found that resonance relaxation can enable PAHs to be aligned up to a degree of $\sim 10~\%$. We believe that this mechanism provides a viable candidate for nanoparticle alignment both in terms of PAHs and silicate nanoparticles in spite of recent controversy about its efficiency (see \citealt{2016ApJ...831...59D}).\footnote{\cite{2016ApJ...831...59D} pointed out that quantization effect may significantly suppress the alignment of nanoparticles because the vibrational energy levels are too broad compared to the intrinsic broadening width, freezing the energy transfer from rotational system to vibrational system. This can lead to negligible polarization of spinning dust emission. However, the treatment of quantum suppression in \cite{2016ApJ...831...59D} is perhaps incomplete because the effect of Raman phonon scattering discussed in \cite{2000ApJ...536L..15L} in relation to the paramagnetic relaxation is disregarded.} Theoretical calculations in \cite{2017ApJ...838..112H} found that, if PAHs are aligned with the magnetic field by paramagnetic resonance mechanism, their mid-IR emission can be polarized to a few percents for the RN conditions. The remaining question is are there any alternative processes that can align PAHs?
   
The history of grain alignment ideas (see \citealt{2003JQSRT..79..881L} for a review) is rich, and various ideas have been tested. In particular, in \cite{Lazarian:1995p3034}, the cross-section alignment mechanism was proposed as a means of aligning disk-like grains. The essence of the mechanism is that the rate of the diffusion of the grain angular momentum can change depending on the cross-section that the planar grain exposes to the flow of atoms. This mechanism relies on the effect that the grain would stay in the position that minimizes its cross section (see also \citealt{Lazarian:1996p6083}). PAH molecules likely exhibit this type of planar geometry. However, instead of atoms which arrival and subsequent evaporation would induce angular momentum diffusion, it is more promising to consider the flow of anisotropic radiation. Therefore, in this paper we explore the cross-sectional alignment of PAHs by anisotropic radiation. 
   
Previous studies for alignment of nanoparticles were carried out for the diffuse cold neutral medium (CNM) (e.g., \citealt{2000ApJ...536L..15L}; \citealt{2013ApJ...779..152H}; \citealt{2014ApJ...790....6H}; \citealt{2016ApJ...831...59D}) where the radiation field can be considered isotropic. For the conditions of reflection nebulae (RNe) and photodissociation regions (PDRs), the unidirectional illumination is present, thus the cross-section alignment mechanism is perhaps important. 

The structure of our paper is as follows. Section \ref{sec:phys} describes the basic physics of PAHs, including magnetic properties and internal relaxation. In Section \ref{sec:FGIR} we present the rotational damping and excitation for anisotropic illumination and calculate the averaged diffusion coefficients over internal thermal fluctuations. In Section \ref{sec:align}, we describe alignment physics and numerical method to compute the degree of PAH alignment. Section \ref{sec:par} presents our obtained results on PAH alignment for the various environment conditions and physical parameters. In Section \ref{sec:polem}, we compute the polarization of spinning dust emission using the resulting degree of alignment for RN, PDR, and CNM. Discussion and summary are presented in Section \ref{sec:concl}.

\section{PAH Physics and Internal Relaxation}\label{sec:phys}
\subsection{PAH geometry}
Let consider a PAH molecule which has an axisymmetric shape with the symmetry axis $\ahat_{1}$ and $\ahat_{2}\ahat_{3}$ lying in the plane. The shape is assumed to be disklike for smallest PAHs and oblate spheroid of axial ratio $s<1$ for large PAHs. The effective size of the molecule is defined as the radius of the sphere having the same volume, and is given by $a=a_{2}s^{1/3}$ where $a_{1}$ and $a_{2}$ are the length of semi-major and semi-minor axes. The transition from disklike to oblate spheroid is set to be at $a=6$\AA~(\citealt{1998ApJ...508..157D}; \citealt{Hoang:2010jy} (hereafter HDL10). Let $I_{\|} $ and $I_{\perp}$ be the moments of inertia along $\ahat_{1}$ and $\ahat_{2}$, and $h_{a}=I_{\|}/I_{\perp}$ (see, e.g., \citealt{2016ApJ...824...18H} for their formulae).

\subsection{Magnetic properties}
Ideal PAHs are expected to have rather low paramagnetic susceptibility due to H nuclear spin \citep{Jones:1967p2924}. However, astrophysical PAHs are likely magnetized thanks to the presence of free radicals, paramagnetic carbon rings, or adsorption of ions (see \citealt{2000ApJ...536L..15L}). 

During the last decade, significant progress has been made in research on magnetism of graphene, providing important insight into magnetism of PAHs. For instance, \cite{2007PhRvB..75l5408Y} suggested that graphene can be magnetized by defects in carbon rings and adsorption of hydrogen atoms to the surface. The vacancy of carbon atoms from the carbon rings creates unpaired electrons, giving rise to the magnetization of graphene \citep{2007PhRvB..75l5408Y}. Also, the adsorption of a hydrogen atom induces magnetic ordering (see \citealt{2004PhRvL..93r7202L}), which is detected in a recent experiment (\citealt{2016Sci...352..437G}). In the ISM, the defects of PAHs can be triggered by bombardment of cosmic rays. All in all, astrophysical PAHs are likely paramagnetic. Let $f_{p}$ be the fraction $f_{p}$ of paramagnetic atoms in the grain, and we take $f_{p}=0.01$ for calculations of PAH magnetic susceptibility, as in previous works (see \citealt{2014ApJ...790....6H} for details).

\subsection{Internal relaxation and thermal fluctuations}\label{sec:Bar}
\cite{Barnett:1915p6353} first pointed out that a rotating paramagnetic body can get magnetized with the magnetic moment along the grain angular velocity. Later, \cite{1976Ap&SS..43..257D} introduced the magnetization via the Barnett effect for dust grains and considered its consequence on grain alignment. 

\cite{1979ApJ...231..404P} realized that the precession of $\bomega$ coupled to $\bmu_{\Bar}$ around the grain symmetry axis $\ahat_{1}$ produces a rotating magnetization component within the grain body coordinates. As a result, the grain rotational energy is gradually dissipated until $\bomega$ becomes aligned with $\ahat_{1}$-- an effect that Purcell termed "Barnett relaxation".

Internal relaxation involves the transfer of grain rotational energy to vibrational modes. Naturally, if the grain has nonzero vibrational energy, energy can also be transferred from the vibrational modes to the rotational energy \citep{Jones:1967p2924}. For an isolated grain, a small amount of energy gained from the vibrational modes can induce fluctuations of the rotational energy $E_{\rot}$ when the grain angular momentum $\bJ$ is conserved (\citealt{1994MNRAS.268..713L}). Over time, the fluctuations in $E_{\rot}$ establish a local thermal equilibrium (LTE). 

The rotational energy is $E_{\rot}=J^{2}\left[1+(h_{a}-1)\sin^{2}\theta\right]/2I_{\|}$ where $\theta$ is the angle between $\ahat_{1}$ and $\bJ$. Thermal fluctuations of $\theta$ can be described by the Boltzmann distribution \citep{1997ApJ...484..230L}:
\bea
f_{\LTE}(\theta,J)= Z{\exp}\left(-\frac{J^{2}}{2I_{\|}k_{\B}T_{d}}
\left[1+(h_{a}-1)\sin^{2}\theta\right]\right),\label{eq:fLTE}
\ena
where $Z$ is the normalization constant such that $\int_{0}^{\pi} f_{\LTE}(\theta,J)\sin\theta d\theta=1$, and $T_{d}$ is the internal alignment temperature above which the vibrational-rotational energy exchange is still effective. 

The temperature of small PAHs undergoes strong fluctuations due to stochastic heating by UV photons (\citealt{Purcell:1976ku}; \citealt{1989ApJ...345..230G}; \citealt{2001ApJ...551..807D}). In this paper, to capture the effect of anisotropic illumination on PAH alignment, $T_{d}$ is assumed to be constant, and a detailed discussion of its effect will be presented in Section \ref{sec:Tfluc}.

\section{Rotational Damping and Excitation by Anisotropic Radiation}\label{sec:FGIR}

\subsection{Rotational Damping and Excitation Coefficients}

The dimensionless damping and excitation coefficients, $F$ and $G$
are defined as 
\bea 
F_{j,r}=-\frac{\tau_{\H,r}}{\omega_{r}}\frac{d\omega_{r}}{dt},\label{eq_F}
\\
G_{j,r}=\frac{\tau_{\H,r}}{2\kB T_{\rm gas}}\frac{I_{r} d\omega_{r}^{2}}{dt},
\label{eq_G}
\ena
where $r=\|,\perp$ denotes the rotation parallel and perpendicular to the grain symmetry axis,
$j$=n, i, p and IR denote collisions of the grain with neutral,
ion, plasma-grain interactions and infrared emission, $\tau_{\rm H,r}$ is the damping time of the grain in a purely HI gas of temperature $T_{\rm gas}$, and $I_{r}$ is the moment of inertia  
along the axis $r$ (see also \citealt{Hoang:2010jy}).

\begin{figure}
\centering
\includegraphics[width=0.3\textwidth]{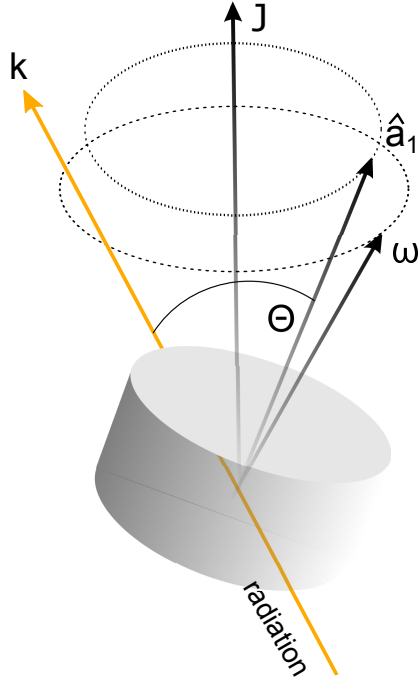}
\caption{A schematic of a disklike PAH molecule with the symmetry axis $\ahat_{1}$, rotating with angular momentum $\bJ$, illuminated by the radiation along the direction $\kv$. The cross section of photon absorption is determined by the angle $\Theta$.}
\label{fig:PAHalign}
\end{figure}

We start by considering photon absorption of a PAH molecule in an isotropic radiation field. Thus, the rate of photon absorption is given by
\bea
\dot{N}_{\abs,0}=\int C_{\rm abs}\frac{cu_{\nu}}{h\nu} d\nu,
\ena
where $u_{\nu}$ is the spectral energy density of the isotropic interstellar radiation in the solar neighborhood with the total energy density $u_{\rm rad}^{\rm M83}=8.63\times 10^{-13} \erg\cm^{-3}$ from \citep{1983A&A...128..212M}, and $C_{\rm abs}$ is the absorption cross-section from PAHs, taken from \cite{2001ApJ...551..807D}. 

Let consider now an anisotropic radiation field of anisotropy degree $\gamma$, with the spectral energy density $u_{\nu}= Uu_{\nu}^{\rm M83}$ and $U$ being the dimensionless scaling factor. Let $\zhat_{k}$ be the anisotropic direction of the radiation that makes an angle $\Theta$ with the PAH symmetry axis $\ahat_{1}$ (see Figure \ref{fig:PAHalign}). To account for the effect of anisotropic illumination, a window $\Psi$ function is introduced (see HDL10):
\bea
\dot{N}_{\rm abs}=U\Psi(\Theta)\dot{N}_{\rm abs,0},
\ena
where
\bea
\Psi(\Theta)=(1-\gamma) + \frac{3}{4}(1+\cos^{2}\Theta)\gamma.
\ena
For an isotropic radiation field, $\gamma=0$, and the function $\Psi(\Theta)$ returns to unity. For an unidirectional radiation ($\gamma=1$), the absorption cross-section at $\Theta=0$ is twice that at $\Psi=\pi/2$ (\citealt{Sironi:2009p5558}; \citealt{2017ApJ...838..112H}).

Let $P_{m,\nu,0}$ be the time-averaged emission power in mode $m$ at frequency $\nu$ due to the absorption of isotropic radiation. Then the total rate of photon emission in the anisotropic case is
\bea
\dot{N}_{m}=\Psi(\Theta)\dot{N}_{m,0}=\Psi(\Theta)\int_{0}^{\infty}\frac{P_{m,\nu,0}}{h\nu}d\nu,\label{eq:dNdt_em}
\ena
where $m=ip$ (in-plane) or $m=oop$ (out-of-plane) emission mode. The averaged wavelength of emission is $\langle \lambda\rangle_{m}=\int_{0}^{\infty}(\frac{\lambda P_{m,\nu,0}}{h\nu}d\nu)/\dot{N}_{m,0}$.

The parallel and perpendicular components of the damping and excitation coefficients induced by photon absorption followed by infrared emission are then given by 
\bea
 F_{\rm IR,\|,0}&=& \frac{3\hbar}{2\pi I_{\|}c} \dot{N}_{\rm ip,0}\langle \lambda \rangle_{ip}\tau_{\H,\|},\\
 F_{\rm  IR,\perp,0}&=&\frac{3\hbar}{2\pi I_{\perp}c} \left[\frac{1}{2}\dot{N}_{\rm ip,0}\langle \lambda \rangle_{ip} +\dot{N}_{\rm oop,0}\langle \lambda_{\rm oop}\right]\tau_{\H,\perp},\label{fir_par}
 \ena
 and 
 \bea
G_{\rm IR,\|,0}&=&\frac{1}{2}\frac{\left[\dot{N}_{\abs,0}+\dot{N}_{\rm ip,0}\right]2\hbar^{2}
}{2I_{\|}\kB T_{\rm gas}}\tau_{\rm H,\|},~~~\label{gir_par}\\
G_{\rm IR,\perp,0}
&=&\frac{1}{2}\frac{\left[\dot{N}_{\rm oop,0}+\frac{1}{2}(\dot{N}_{\abs,0}+\dot{N}_{ip,0})\right]2\hbar^{2}}
{2I_{\perp}\kB T_{\rm gas}}\tau_{\rm H,\perp},
~~~~\label{gir_per} 
\ena 
where the recoil of angular momentum $L=\hbar \sqrt{j(j+1)}=\sqrt{2}\hbar$ has been used.

For an anisotropic radiation field, the damping and excitation coefficients become
\bea
F_{\IR,j}(\Theta)= U \Psi(\Theta)F_{\IR,j,0},\label{eq:FIR}\\
G_{\IR,j}(\Theta)= U \Psi(\Theta)G_{\IR,j,0},\label{eq:GIR}
\ena
where $j=\|,\perp$. The damping and excitation coefficients are thus maximum for $\Theta=0$ and minimum for $\Theta=\pi/2$.

The total IR coefficients are given by $F_{\IR,\rm neu}f_{0}+ (1-f_{0})F_{\IR,\rm ion}$ where $\rm neu$ and $\rm ion$ denote neutral PAH and ionized PAH (both positive and negative charge states). Here we assume neutral PAHs ($f_{0}=1$) in calculations of $F_{\IR}$ and $G_{\IR}$.

To compute $F_{\IR,j}$ and $G_{\IR,j}$ induced by anisotropic radiation with Equations (\ref{eq:FIR}) and (\ref{eq:GIR}), we take $\langle \lambda\rangle,\dot{N}_{ip,0},\dot{N}_{oop,0}$, and $\dot{N}_{\abs,0}$ computed in HDL10. For circumstellar regions or reflection nebula, we assume that the spectrum of radiation is similar to that of the standard local interstellar radiation field (ISRF; \citealt{1983A&A...128..212M}), but the strength of the radiation is described by the parameter $U$. Other $F$ and $G$ coefficients are calculated as in HDL10, where the charge distribution of nanoparticles arising from collisional charging and photoelectric effect is computed as in \cite{2012ApJ...761...96H}.

\subsection{Averaging the damping and diffusion coefficients over thermal fluctuations}
For a fixed angular momentum, the $F_{\rm IR}(\Theta)$ and $G_{\rm IR}(\Theta)$ coefficients fluctuate due to the fluctuations of the angle $\theta$ between $\ahat_{1}$ and $\bJ$. To consider the effect of the coefficients on the alignment, we first need to transform the coefficients to the lab system, and then average it over the thermal fluctuations.

\subsubsection{Transformation}
\begin{figure*}
\centering
\includegraphics[width=0.8\textwidth]{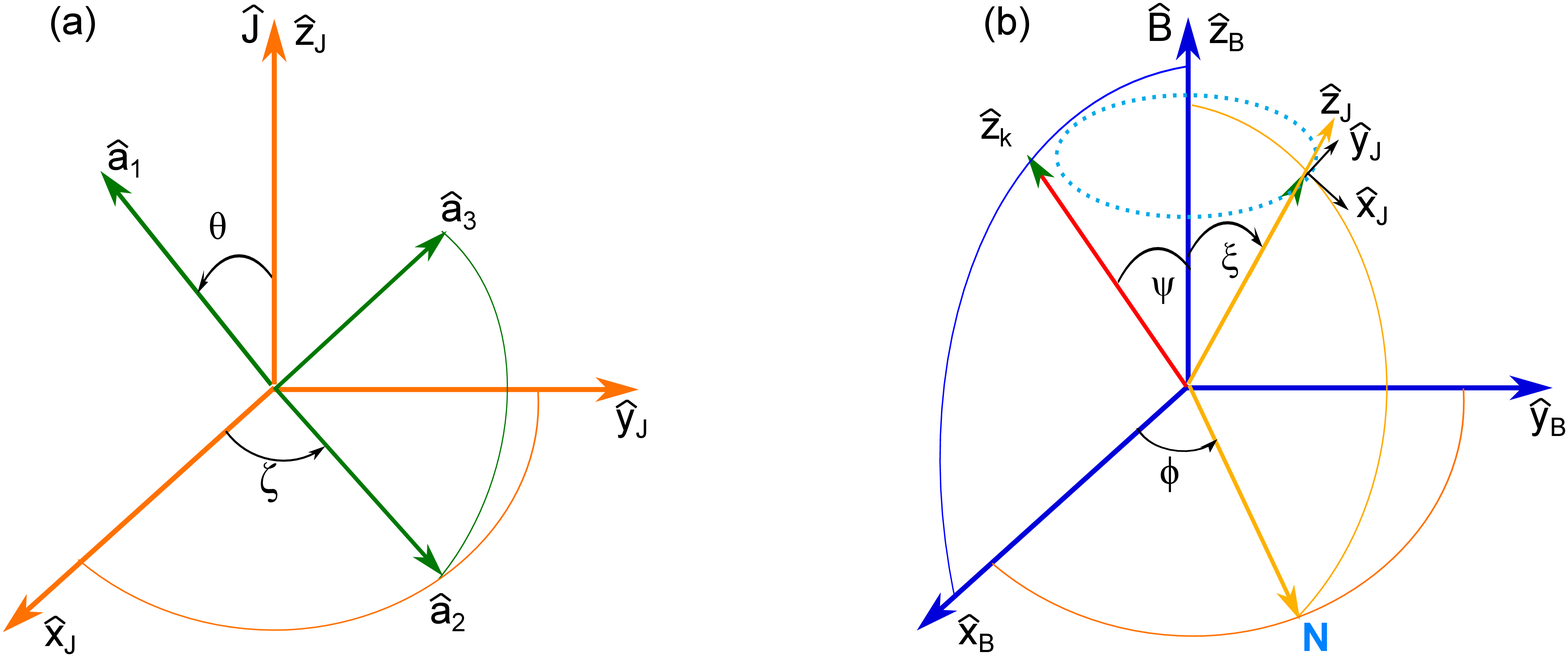}
\caption{Left panel: orientation of an axisymmetric grain described by three principal axes $\ahat_{1}\ahat_{2}\ahat_{3}$ in the coordinate system defined by $\bJ$. The axis $\ahat_{2}$ is chosen along the line of node. Right panel: orientation of $\bJ$ in the coordinate system defined by the magnetic field $\Bv\| \zhat_{B}$. $\zhat_{J}\| \bJ$, $\xhat_{J}\equiv \hat{\xi}$, and $\yhat_{J}\equiv \hat{\phi}$.}
\label{fig:RFs}
\end{figure*}

To study the alignment of the grain angular momentum $\bJ$ with the ambient magnetic field $\Bv$, we consider an inertial lab coordinate system denoted by unit vectors $\xhat_{B}\yhat_{B}\zhat_{B}$ such that $\zhat_{B}\|\Bv$ (see Figure \ref{fig:RFs}, right panel).  

The coefficients $F_{\rm IR}(\Theta)$ and $G_{\rm IR}(\Theta)$ depend on the instantaneous orientation of the PAH molecule in the lab system. Thus, first, we have to transform these coefficients from the grain body to the lab reference frame.

For the diffusion coefficients, $B_{ij}^{b}=\langle {\Delta J_{i}^{b}\Delta J_{j}^{b}}/{\Delta t}\rangle$ with $B_{ij}^{b}=0$ for $i\ne j$. These diffusion coefficients are related to the dimensionless excitation coefficients $G_{j}$ as follows:
\bea
B_{j,zz}^{b}=B_{j,\|}^{b}=\frac{2I_{\|}\kB T_{\rm gas}}{\tau_{\rm H,\|}}G_{j,\|},\\
B_{j,xx}^{b}=B_{yy}^{b}=B_{j,\perp}^{b}=\frac{2I_{\perp}\kB T_{\rm gas}}{\tau_{\rm H,\perp}}G_{j,\perp}.
\ena
where $j=n, ion, p, IR$ denotes the various interaction processes.

The transformation of the diffusion coefficients from the body frame to the lab system is carried out as in HDL10 (see also \citealt{1997MNRAS.288..609L}), and we obtain $A_{j,i}$ and $B_{j,ii}$ in the lab system for all processes $j$:
\bea
B_{j,ii} = B_{j,\|}^{b}g_{j,\|} + B_{j,\perp}^{b}g_{j,\perp},\label{eq:Btrans}
\ena
where $g_{i,\|}$ and $g_{i,\perp}$ are the transformation functions  which depends on $j$ (see Appendix \ref{apdx:transform} for details).

In dimensionless units of $J'=J/I_{\|}\omega_{T,\|}$ and $t'=t/\tau_{\H,\|}$,  the coefficients have denoted with prime. Using Equation (\ref{eq:Btrans}), the dimensionless coefficients for IR emission at a given orientation of $\bJ$ described by $(\xi,\phi,\psi)$ can be written as
\bea
A'_{\IR,i}(\bJ,\theta,\zeta)&=&F_{\IR,\|}(\Theta)\left(\cos^{2}\theta +\gamma_{\rm H}\sin^{2}\theta\right),\\
B'_{\IR,ii}(\bJ,\theta,\zeta) &=& G_{\IR,\|}(\Theta)g_{i,\|} + G_{\IR,\perp}(\Theta)g_{i,\perp}\left(\frac{\epsilon_{\rm H}}{h_{a}}\right),\label{eq:BIR_lab}
\ena
where $i=x,z$, $\epsilon_{\H}=\tau_{\H,\|}/\tau_{\H,\perp}$, $\gamma_{\rm H}=(F_{\rm IR,\perp}/F_{\rm IR,\|})\epsilon_{\rm H}$, and the angle $\theta,\zeta$ describes the orientation of $\ahat_{1}$ with $\bJ$ (see Appendix \ref{apdx:transform}).

The angle $\Theta$ between the illumination direction and the grain symmetry axis is determined by 
\bea
\cos\Theta = \ahat_{1}.\zhat_{k},
\ena
which is a function of five angles $(\zeta, \theta,\phi,\xi,\psi)$ (see Eqs (\ref{eq:a1_zk})-(\ref{eq:yj_zk})).

\subsubsection{Averaging over thermal fluctuations}
The alignment of PAHs with the magnetic field due to magnetic relaxation occurs on a timescale much longer than the rotation period of the PAH around its symmetry axis, the internal relaxation. Therefore, we can average the diffusion coefficients over these fast processes.

First, we average $A'_{\rm IR}(\bJ,\theta,\zeta)$ and $B'_{\rm IR,ii}(\bJ,\theta,\zeta)$ over the rotation angle $\zeta$ of the grain around $\ahat_{1}$. Then, we average the obtained coefficients over the internal thermal fluctuations. The resulting averaged diffusion coefficients are equal to
\bea
\bar{A'}_{\rm IR,i} (J,\xi,\phi,\psi)&=& \frac{1}{2\pi}\int_{0}^{\pi} d\theta\int_{0}^{2\pi}d\zeta A'_{\rm IR,i}(\bJ,\theta,\zeta) f_{\rm LTE},~~~~\\
\bar{B'}_{\rm IR,ii}(J,\xi,\phi,\psi)&=& \frac{1}{2\pi}\int_{0}^{\pi}d\theta \int_{0}^{2\pi} d\zeta B'_{\rm IR,ii}(\bJ,\theta,\zeta)f_{\rm LTE},~~~~\label{eq:Bavg}
\ena
where the integration over $\zeta$ can be separated from $\theta$ due to the much faster rotation around $\ahat_{1}$.

Finally, we average over the Larmor precession:
\bea
\langle B'_{\rm IR,ii}\rangle(J,\xi,\psi) =\frac{1}{2\pi}\int d\phi \bar{B'}_{\rm IR,ii}(J,\xi,\phi,\psi),
\ena
where $ii=xx,yy,zz$. The Larmor precession averaging is straightforward by replacing $\cos^{2}\phi$ with $\langle \cos^{2}\phi\rangle =\int_{0}^{2\pi} d\phi \cos^{2}\phi/2\pi=1/2$ and $\langle \sin^{2}\phi\rangle=1/2$.

\subsubsection{Numerical results}
For each radiation direction $\psi$, we compute the averaged coefficients $\langle A'_{\IR}\rangle$ and $\langle B'_{\IR}\rangle$ as functions of $J'=0.1-10$ and $\xi=0-\pi$. Then, we will use the tabulated data $\langle A'_{\IR}\rangle(J,\xi)$ and $\langle B'_{\IR}\rangle(J,\xi)$ for solving the equations of motion in Section \ref{sec:LE}. 

\begin{figure*}
\centering
\includegraphics[width=0.4\textwidth]{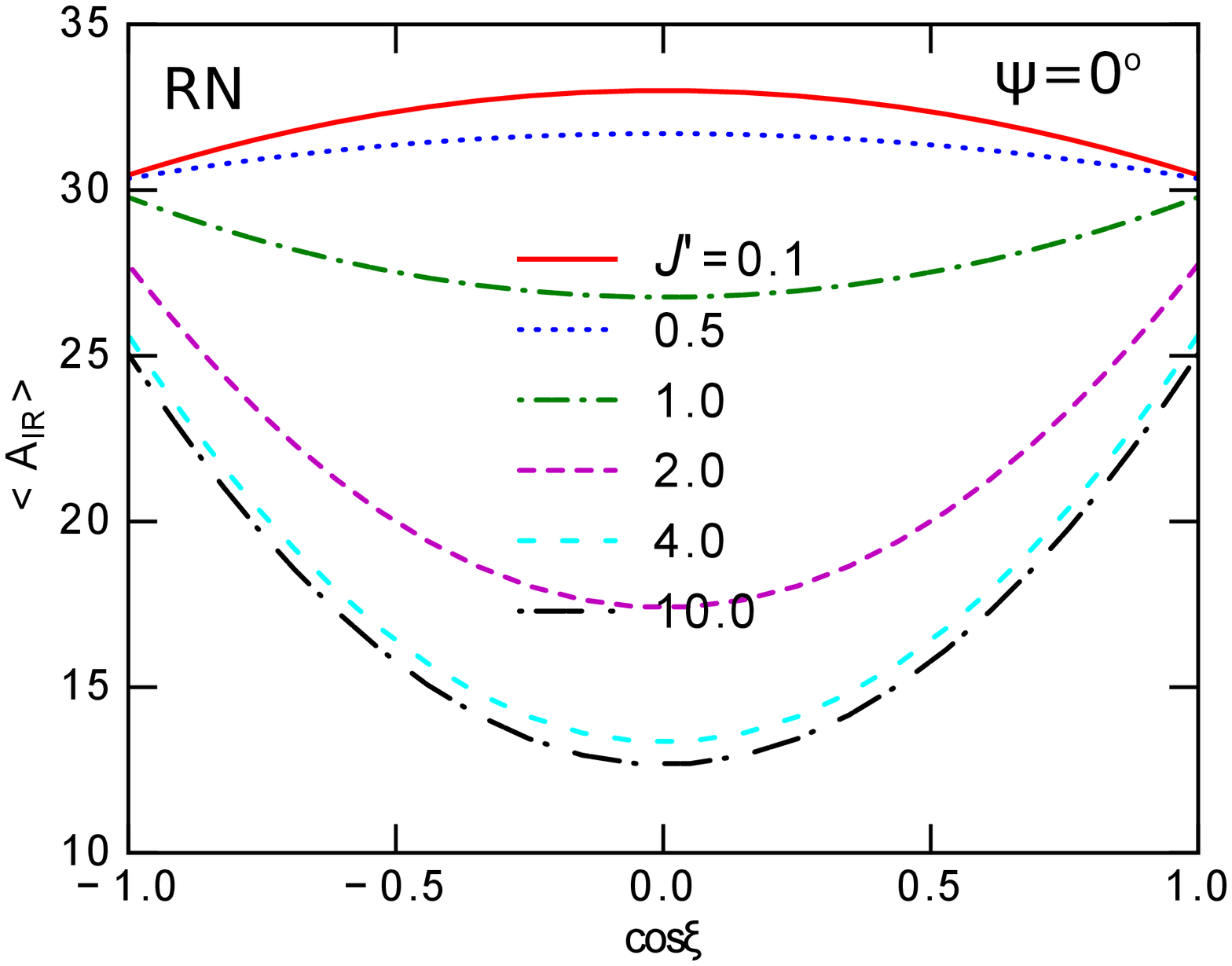}
\includegraphics[width=0.4\textwidth]{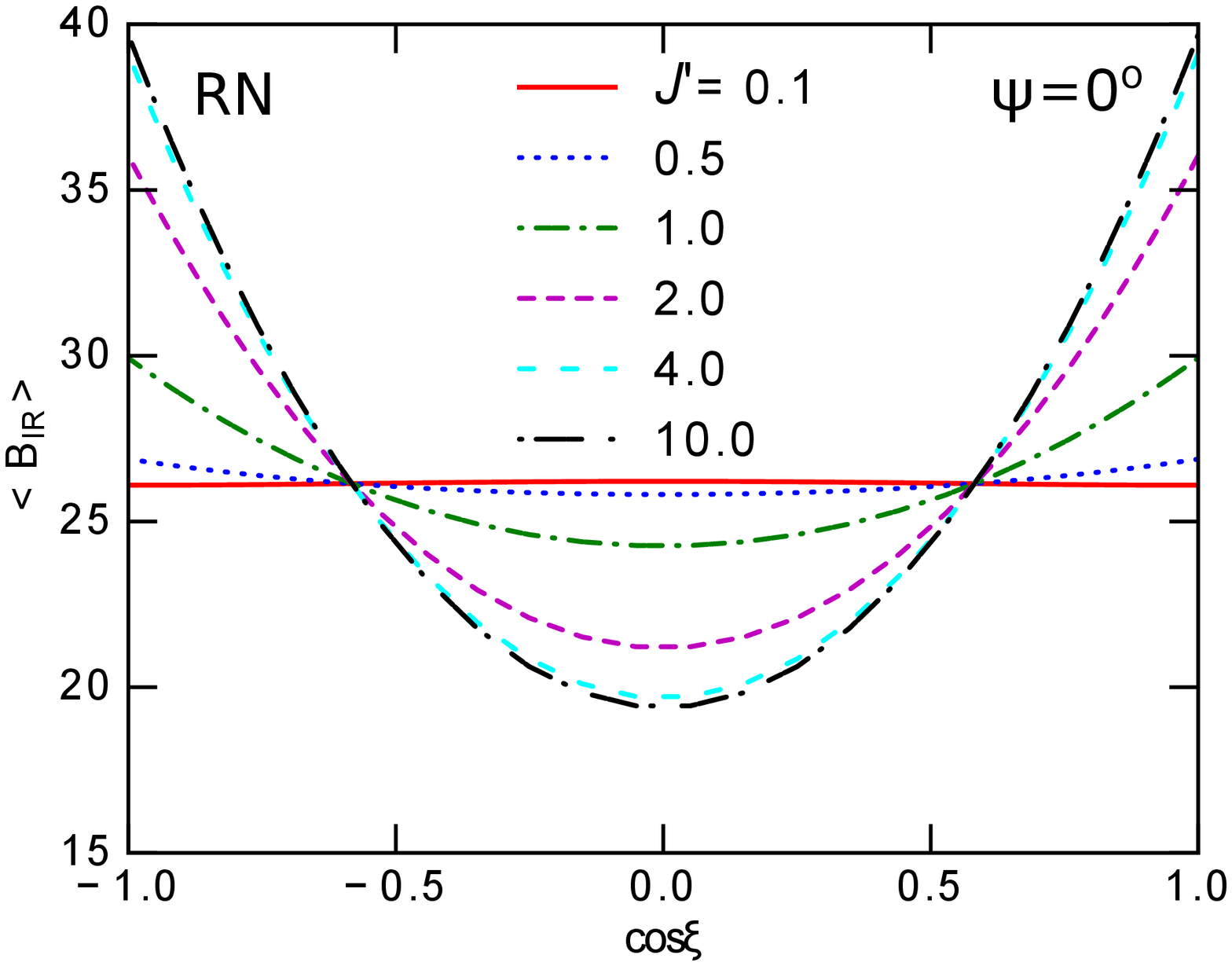}
\includegraphics[width=0.4\textwidth]{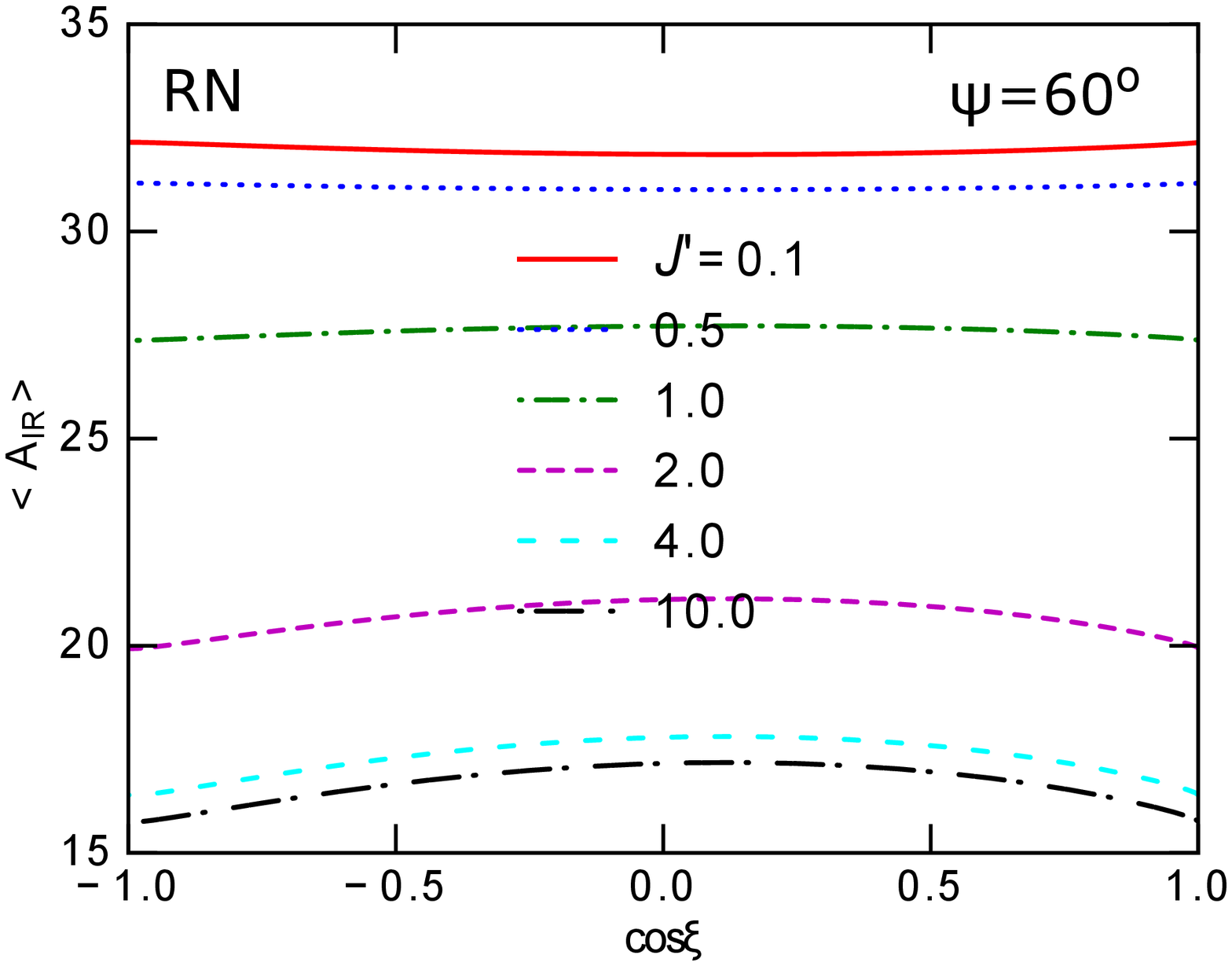}
\includegraphics[width=0.4\textwidth]{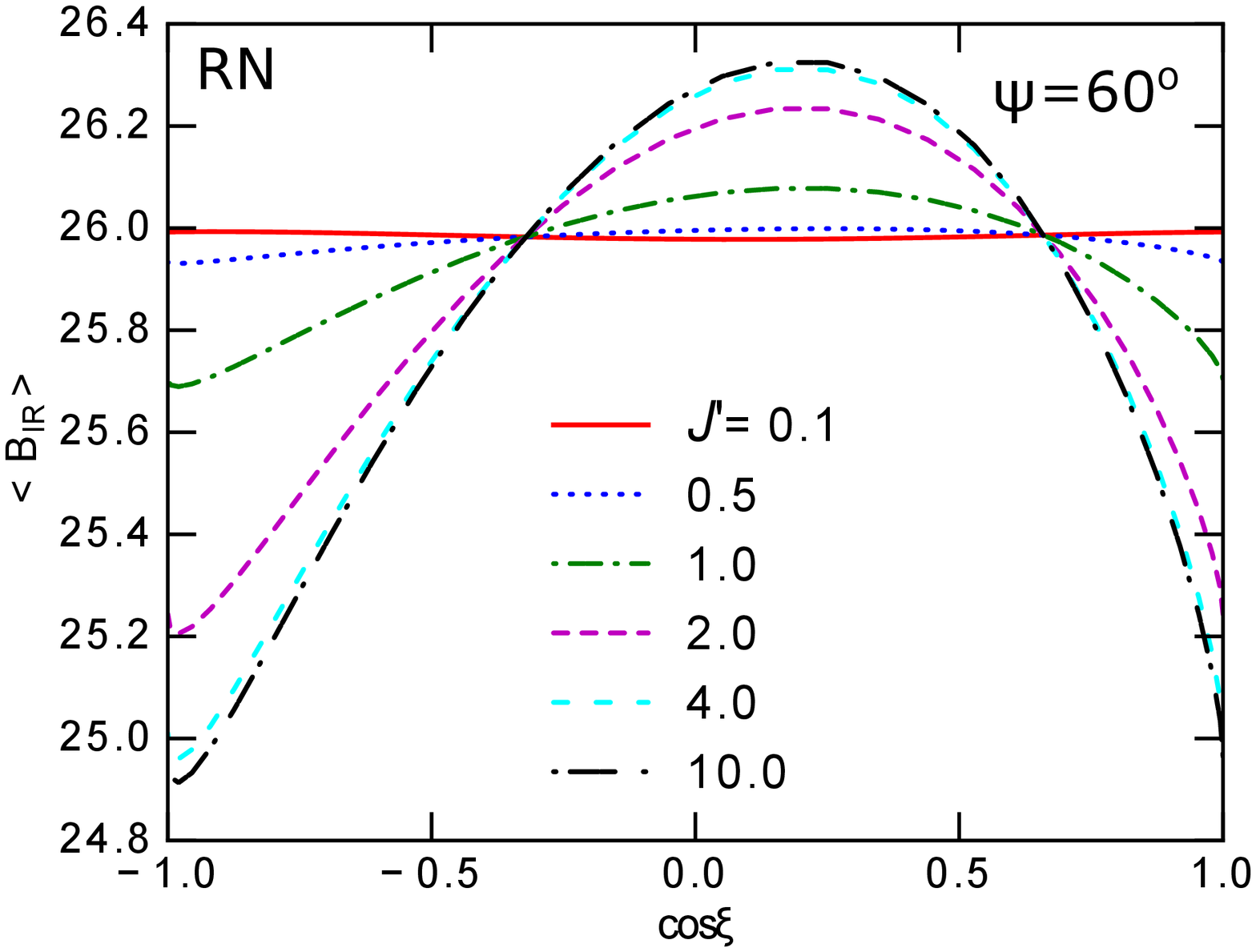}
\caption{Averaged IR damping coefficient (left panel) and excitation coefficient (right panel) vs. $\cos\xi$ for the different values of grain angular momentum $J'$ in the RN conditions. Two illumination angles $\psi=0^{\circ}$ (upper panels) and $\psi=60^{\circ}$ (lower panels) are considered.}
\label{fig:ABIR_psi0}
\end{figure*}

Figure \ref{fig:ABIR_psi0} (upper panels) shows $\langle A'_{\IR}\rangle(J,\xi)$ and $\langle B_{\IR}\rangle(J,\xi)$ as a function of the angle $\xi$ for several values of $J$ and $\psi=0^{\circ}$. For the case of subthermal rotation (i.e., $J'\le 0.5$),  $\langle A'_{\IR}\rangle$ is essentially similar for the different $\xi$ because the strong thermal fluctuations result in nearly isotropic distribution of the PAH orientation in the space. For $J'\gg 1$, the thermal fluctuations are very weak due to suprathermal rotation, thus the PAH is spinning with $\ahat_{1}$ aligned with $\bJ$. As a result, $\langle A'_{\rm IR}\rangle$ is maximum at $\xi=0,\pi$, i.e., when the PAH plane is perpendicular to the incident radiation such that the photon absorption cross-section is maximum. The coefficients are minimum at $\xi=\pi/2$, i.e., depending on the anisotropy direction. For $\psi=60^{\circ}$, the maximum coefficient occur at $\cos\xi=\pi/2$, shifted from $\cos\xi=\pm 1$.

\section{PAH alignment by anisotropic radiation field}\label{sec:align}

\subsection{Resonance Paramagnetic Relaxation}
\cite{1951ApJ...114..206D} suggested that a paramagnetic grain rotating with angular velocity $\bomega$ in an external magnetic field $\Bv$ experiences paramagnetic relaxation due to the lag of magnetization, which dissipates the grain rotational energy into heat. This results in the gradual alignment of $\bomega$ and $\bJ$ with the magnetic field until the rotational energy is minimum.

For ultrasmall grains, such as PAHs, the classical Davis-Greenstein relaxation is suppressed because the rotation time is shorter than the electron-electron spin relaxation time $\tau_{2}$ (see \citealt{2014ApJ...790....6H}). Yet, such nanoparticles can be partially aligned by resonance paramagnetic relaxation that originates from the splitting of the rotational energy \citep{2000ApJ...536L..15L}. Numerical calculations in \cite{2014ApJ...790....6H} showed that PAHs can be aligned by resonance paramagnetic relaxation with the degree $Q_{J}\sim 0.05-0.15$ depending on $T_{0}$ and the magnetic field strength. Such a considerable degree of external alignment will produce anisotropy distribution of the angular momentum in the space, enhancing the polarization of PAH emission features. \cite{2016ApJ...821...91H} extended alignment calculations for iron nanoparticles and nanosilicates.

\subsection{Numerical Method}\label{sec:LE}
To study the alignment of PAHs with the magnetic field, we solve the Langevin equations (LEs): 
\bea
dJ_{i}=A_{i}dt+\sqrt{B_{ii}}dW_{i}\mbox{~for~} i=~1,~2,~3,\label{eq:dJ_dt}
\ena
where $dW_{i}$ are the random variables drawn from a normal distribution with zero mean and variance $\langle dW_{i}^{2}\rangle=dt$, and $A_{i}=\langle {\Delta J_{i}}/{\Delta t}\rangle$ and $B_{ii}=\langle \left({\Delta J_{i}}\right)^{2}/{\Delta t}\rangle$ are the damping and diffusion coefficients defined in the lab $\ehat_{1}\ehat_{2}\ehat_{3}$ system, respectively. Here, $\ehat_{1}\equiv \zhat_{B}$, $\ehat_{2}\equiv \xhat_{B}$, and $\ehat_{3}=\yhat_{B}$ (see Figure \ref{fig:RFs}, right panel).

To incorporate paramagnetic relaxation, we add a damping term $-{J_{2,3}}/{\tau_{\rm mag}}$ to $A_{2,3}$ and an excitation term $B_{{\rm mag},22}=B_{{\rm mag},33}$ to $B_{22}$ and $B_{33}$, respectively. In the dimensionless units of $J'$ and $t'$, Equation (\ref{eq:dJ_dt}) becomes 
\bea
dJ'_{i}=A'_{i}dt'+\sqrt{B'_{ii}}dw'_{i} \mbox{~for~} i= 1,~2,~3,\label{eq:dJp_dt}
\ena
where $\langle dw_{i}^{'2}\rangle=dt'$ and
\bea
A'_{i}&=&-{J'_{i}}\left[\frac{1}{\tau'_{\gas,{\eff}}} +\delta_{m}(1-\delta_{1i})\right] -\frac{2}{3}\frac{J_{i}^{'3}}
{\tau'_{\ed,{\eff}}},\label{eq:Ai}~~~~\\
B'_{ii}&=&\frac{B_{ii}}{2I_{\|}\kB T_{\gas}}\tau_{\H,\|}+\frac{T_{\d}}{T_{\gas}}\delta_{\rm mag}(1-\delta_{1i}).\label{eq:Bii}
\ena
Above, $\delta_{1i}=1$ for $i=1$ and $\delta_{1i}=0$ for $i\ne 1$, and
\bea
\tau'_{\gas,{\eff}}= \frac{\tau_{\gas,{\eff}}}{\tau_{\H,\|}},~\tau'_{\ed,{\eff}}&=&\frac{\tau_{\ed,{\eff}}}{\tau_{\H,\|}},~~~
\ena
where $\tau_{\gas,{\eff}}$ and $\tau_{\ed,{\eff}}$ are the effective damping times due to dust-gas interactions and electric dipole emission (see Eq. E4 in HDL10). 

The diffusion coefficients for gas-grain interaction (including neutral, ion, and plasma drag) is independent on the grain orientation and its angular momentum, whereas the diffusion coefficients for IR emission depend on $\bJ$. So, we can write
\bea
A'_{i} &=& \sum_{j=n,ion,p}A'_{j} + \langle A'_{\IR,i}\rangle(J,\xi,\psi),\\
B'_{ij} &=& \sum_{j=n,ion,p}B'_{ij,j} + \langle B'_{\IR,ij}\rangle(J,\xi,\psi),
\ena
where $\langle A'_{\rm IR}\rangle(J,\xi,\psi)$ and $\langle B'_{\rm IR}\rangle(J,\xi,\psi)$ are tabulated in the preceding section.

To solve the Langevin equation (\ref{eq:dJp_dt}) numerically, we use the second-order integrator  \citep{VandenEijnden:2006gp}. Thus, the angular momentum component $j_{i}\equiv J'_{i}$ at iterative step $n+1$ is evaluated as follows:
\bea
{j}_{i;n+1} &=& j_{i;n}  -  \gamma_{i}{j}_{i;n}h+\sqrt{h}\sigma_{ii}{\zeta}_{n}- \gamma_{i}\mathcal{A}_{i;n}-\gamma_{\ed}\mathcal{B}_{i;n},~~~
\ena
where $h$ is the timestep, $\gamma_{i}=1/\tau'_{\gas,{\eff}} +\delta_{m}(1-\delta_{zi})$, $\gamma_{\ed}=2/(3\tau'_{\ed,\eff})$, $\sigma_{ii}=\sqrt{B'_{ii}}$, and
\bea
\mathcal{A}_{i;n}&=& -\frac{h^{2}}{2} \gamma_{i} j_{i;n}+\sigma_{ii} h^{3/2}g(\zeta_{n},\eta_{n})-\gamma_{\ed}j_{i;n}^{3}\frac{h^{2}}{2},\\
\mathcal{B}_{i;n}&=&j_{i;n}^{3}h - 3\gamma_{i}j_{i;n}^{3}\frac{h^{2}}{2}-\frac{3j_{i;n}^{5}\gamma_{\ed}h^{2}}{2}+3j_{i;n}^{2}\sigma_{ii} h^{3/2}g(\zeta_{n},\eta_{n}),\nonumber
\ena
with ${\eta}_{n}$ and ${\zeta}_{n}$ being independent Gaussian variables with zero mean and unit variance and $g(\zeta_{n},\eta_{n})= (\zeta_{n} + \eta_{n}/\sqrt{3})/2$ (see \citealt{2016ApJ...821...91H} for details).

The timestep $h$ is chosen by $h= 0.01\min[1/F_{\tot,\|}, 1/G_{\tot,\|}, \tau_{\rm ed,\|}/\tau_{\H,\|},1/\delta_{m}]$.  As usual, the initial grain angular momentum is assumed to have random orientation in the space and magnitude $J=I_{\|}\omega_{T}$ (i.e., $j=1$).

\section{Parameter Space Study for PAH alignment}\label{sec:par}
\subsection{Definition of Alignment Degree}
Let $Q_{X}=\langle G_{X}\rangle $ with $G_{X}=\left(3\cos^{2}\theta-1\right)/2$ be the degree of internal alignment of the grain symmetry axis $\ahat_{1}$ with $\bJ$, and let $Q_{J}=\langle G_{J}\rangle$ with $G_{J}=\left(3\cos^{2}\xi-1\right)/2$ be the degree of external alignment of $\bJ$ with $\Bv$. Here the angle brackets denote the average over the ensemble of grains. The net degree of alignment of $\ahat_{1}$ with $\Bv$, namely the Rayleigh reduction factor, is defined as $R = \langle G_{X}G_{J}\rangle$ \citep{Greenberg:1968p6020}.

We calculate the degree of PAH alignment by solving the Langevin equations. The LE is numerical solved for an integration time $T=10^{3}t_{\gas}$ and $N$ steps, which ensures that $T$ is much larger than the longest dynamical timescale to provide good statistical calculations of the degrees of grain alignment. The degree of alignment is then calculated by replacing the ensemble average with the time averaging as follows: 
\bea
 Q_{J}&\equiv& \sum_{n=1}^{N} \frac{G_{J}(\cos^{2}\xi_{n})}{N},\\
 Q_{X}&=&\frac{1}{N}\sum_{n=1}^{N}\frac{\bar{G}_{X}}{N},\\
 R&=& \sum_{n=1}^{N} \frac{G_{J}(\cos^{2}\xi_{n})\bar{G}_{X})}{N},
 \ena
where $\bar{G}_{X}=\int_{0}^{\pi}G_{X}(\cos^{2}\theta)f_{\rm LTE}(\theta, J_{n})d\theta$. 

The grain rotational temperature can be calculated as $T_{\rot}=\langle J^{2}\rangle/k_{\B}I_{\|}$ where $\langle J'^{2}\rangle =\sum_{i=1}^{N}J'^{2}/N$. 

\subsection{Numerical Results}
\subsubsection{Unidirectional Radiation Field}
We first consider the conditions of RN, where the illumination is unidirectional with anisotropy degree $\gamma_{\rad}=1$. We perform calculations by varying four parameters, the hydrogen ionization fraction $x_{\rm H}$, the magnetic field $B$, the radiation intensity $U$, and the illumination direction $\psi$. For each varying parameter, the rest of physical parameters are assumed to be the typical values of the RN, as shown in Table \ref{tab:ISM}.

\begin{figure*}
\centering
\includegraphics[width=0.9\textwidth]{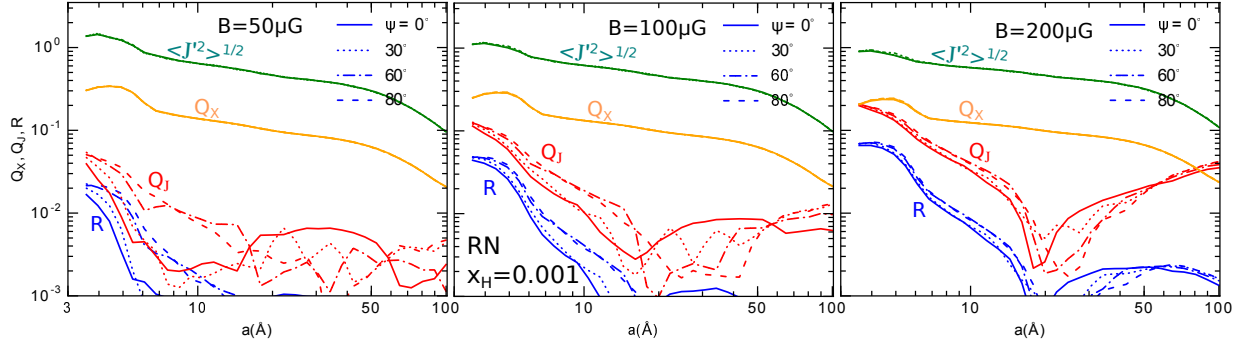}
\caption{Degrees of PAH alignment in the RN conditions vs. the grain size for the different illumination angles ($\psi$) and the hydrogen ionization fraction $x_{\rm H}=0.001$. Left, middle, and right panels show the results for $B=50, 100, 200~\mu$G, respectively.}
\label{fig:RQJ_RN1}
\end{figure*}

Figure \ref{fig:RQJ_RN1} shows the degrees of alignment, $Q_{X},Q_{J},R$ and $\langle J'^{2}\rangle^{1/2}$ for the various illumination direction $\psi$, for $B=50~\mu$G (left panel), $100~\mu$G (middle panel) and $200~\mu$G (right panel).

The degree of internal alignment $Q_{X}$ and $\langle J'^{2}\rangle^{1/2}$ do not change significantly with $\psi$, whereas $Q_{J}$ and $R$ of smallest PAHs ($a<20$~\AA) tend to increase with increasing $\psi$.\footnote{Note that the numerical errors in $R$ obtained with the Langevin simulations is $\sim 0.001$ (\citealt{1999MNRAS.305..615R}; \citealt{2014ApJ...790....6H}), thus, we are not discussing the results for $a>20$~\AA.}
This is expected because for $\psi=0^{\circ}$, the maximum cross-section of PAH absorption corresponds to the PAH plane perpendicular to the magnetic field. Thus, this realization induces strongest damping and randomization of PAH orientation, reducing the degree of alignment. When the magnetic field increases, the degree of external alignment increases thanks to stronger magnetic relaxation, but the increase in alignment degree with $\psi$ is reduced.

\begin{figure*}
\centering
\includegraphics[width=0.9\textwidth]{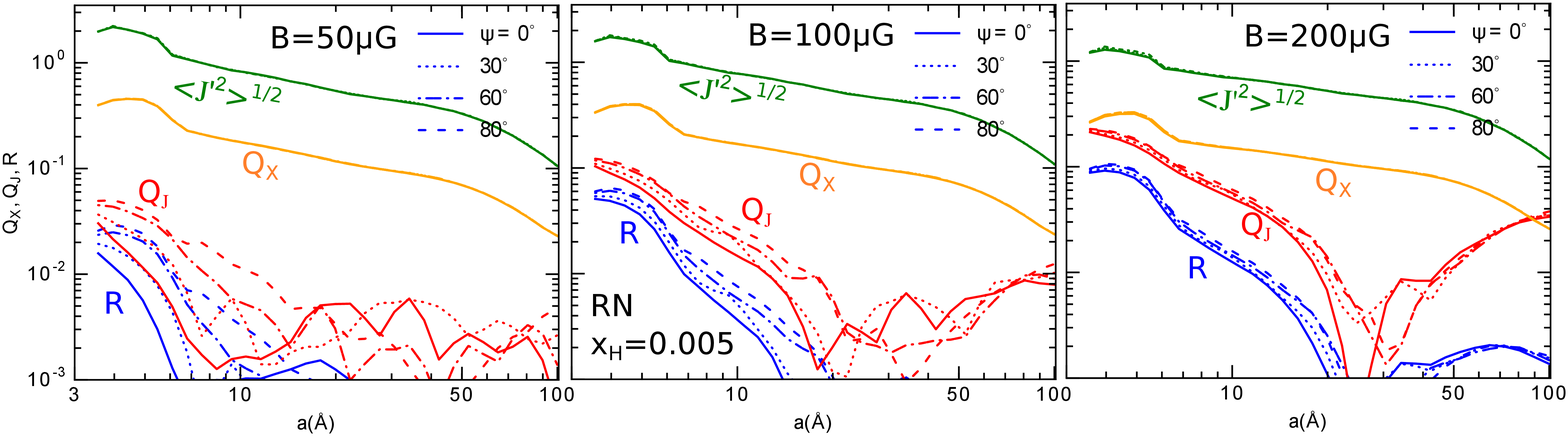}
\includegraphics[width=0.9\textwidth]{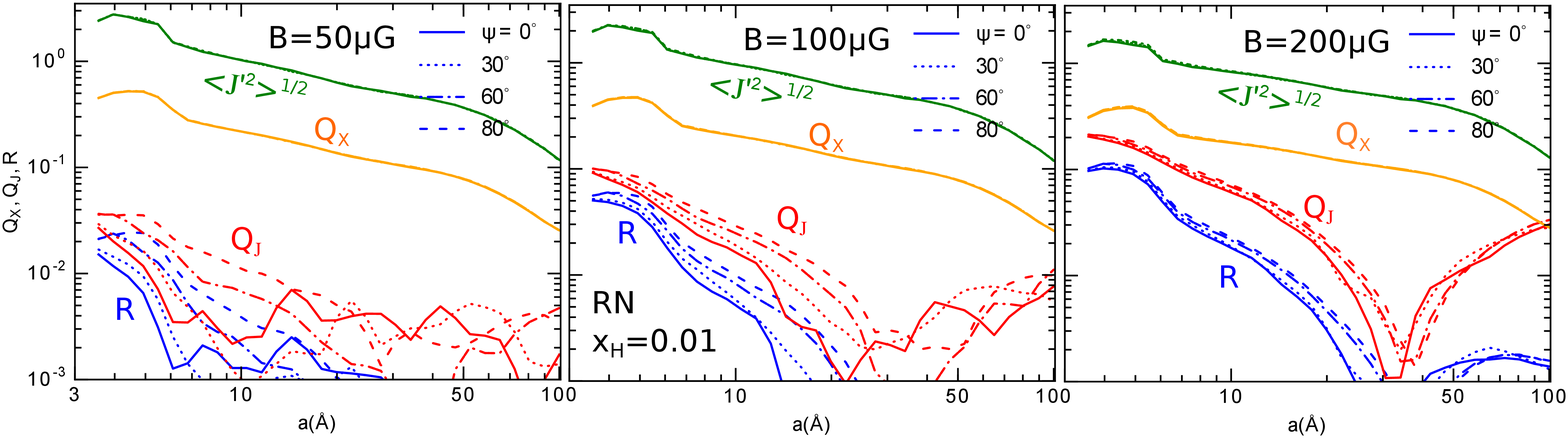}
\caption{Same as Figure \ref{fig:RQJ_RN1}, but for $x_{\rm H}=0.005$ (upper panels) and $0.01$ (lower panels).}
\label{fig:RQJ_RN2}
\end{figure*}

Figure \ref{fig:RQJ_RN2} shows the results for the higher ionization fraction $x_{\rm H}=0.005$ (upper panels) and $0.01$ (lower panels). For the same magnetic field strength, we see the increase of $\langle J'^{2}\rangle^{1/2}$ and then the degrees of alignment with increasing $x_{\rm H}$, which is originated from ion collisions. 

\begin{table}
\caption{Typical parameters for the idealized ISM phases}\label{tab:ISM}
\begin{tabular}{l l l l} \hline\hline\\
{\it Parameters} & {CNM} & RN & PDR\\[1mm]
\hline\\
$n_{\rm H}$(cm$^{-3}$) &30 &$10^{3}$  &$10^{5}$\\[1mm]
$T_{\rm gas}$(K)& 100 &100 & 1000 \\[1mm]
$T_{\rm d}$(K)& 20 & 40 & 80 \\[1mm]
$U$ &1 &$1000$ & $30000$\\[1mm]
$x_{\rm H}\equiv n(\H^{+})/n_{\H}$ &0.0012 &$0.001$ & $0.0001$ \\[1mm]
$x_{\rm M}\equiv n(\rm M^{+})/n_{\H}$ &$0.0003$ & $0.0002$ & $0.0002$\\[1mm]
$y=2n(\H_{2})/n_{\H}$ & {$0.0$} & 0.5 & 0.5\\[1mm]
$B(\mu \rm G)$&{$10$}& $100$ & 100\\[1mm]
\\[1mm]
\hline\hline\\
\end{tabular}
\end{table}

\begin{figure*}
\centering
\includegraphics[width=0.7\textwidth]{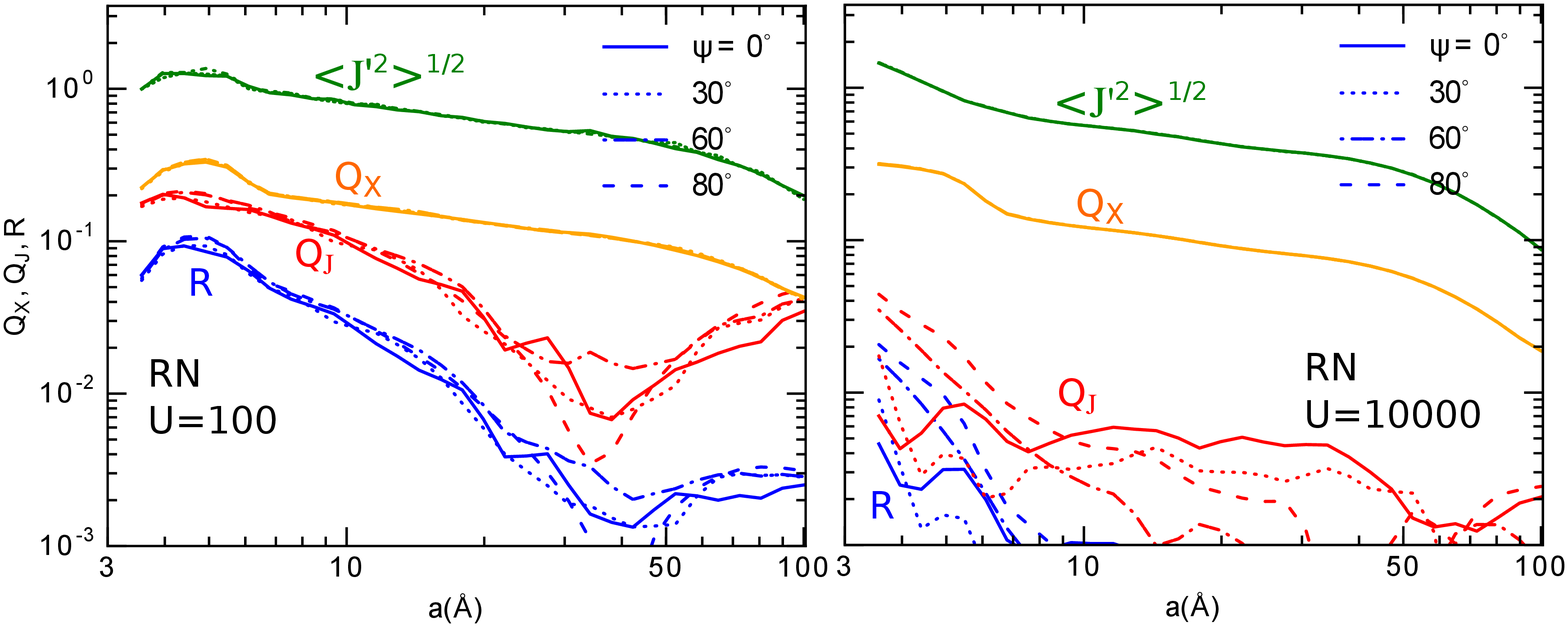}
\caption{Same as Figure \ref{fig:RQJ_RN1} but for a weak radiation field ($U=10^{2}$) and very strong radiation field ($U=10^{4}$).}
\label{fig:RQJ_RN_U}
\end{figure*}

To see the dependence of alignment degree on the radiation intensity, we show in Figure \ref{fig:RQJ_RN_U} the alignment degree for two different radiation fields. The stronger radiation field results in a much lower degree of alignment due to more efficient damping by IR emission (see the right panel). However, the effect of anisotropic illumination is much stronger for this case, leading to a significant increase of $Q_{J}$ and $R$ with $\psi$ for small PAHs below 10~\AA.

We note the insensitive of rotational temperature on the illumination direction $\psi$. This reveals that the enhanced alignment with $\psi$ is mainly due to the differential cross-section, not due to paramagnetic relaxation.

\subsubsection{The diffuse interstellar medium: CNM}
\begin{figure}
\includegraphics[width=0.45\textwidth]{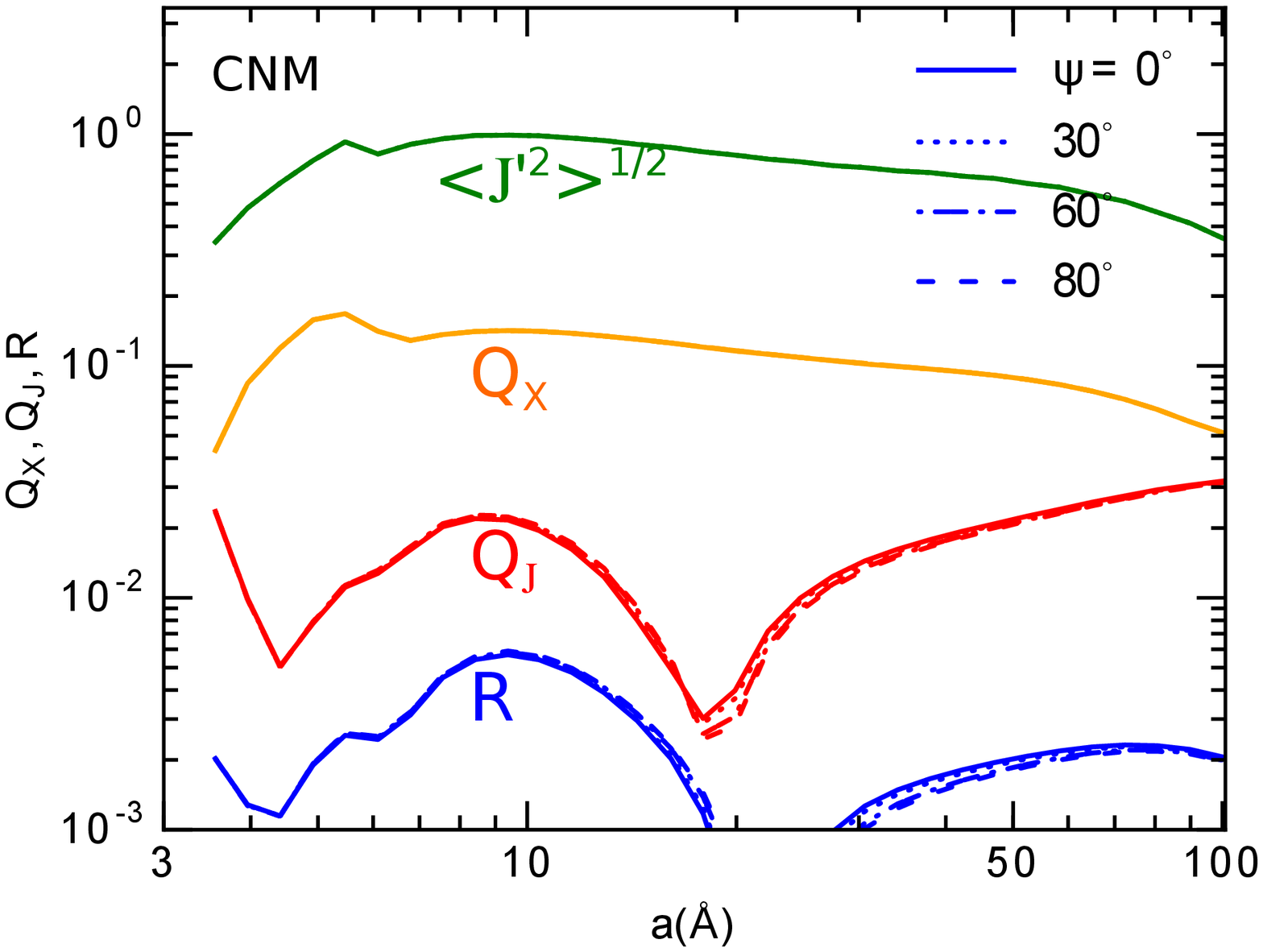}
\caption{Degrees of PAH alignment in the CNM with an anisotropy degree $\gamma=0.1$ and $B=10~\mu$G.}
\label{fig:RQJ_CNM}
\end{figure}

For completeness, Figure \ref{fig:RQJ_CNM} shows the results for the diffuse interstellar medium (CNM) with the typical anisotropy degree of radiation $\gamma=0.1$. A sharp rise at $a \sim 4.5$\AA~ is a numerical artifact, which is well below the numerical accuracy achieved by Langevin equation simulations (see, e.g., \citealt{1999MNRAS.305..615R}). The dependence of PAH alignment on the illumination angle is negligible, as expected from the low degree of anisotropy.

\section{Polarization of spinning dust emission}\label{sec:polem}
\subsection{Spinning Dust Model}

We calculate the spinning dust emissivity and polarized emissivity per H is calculated as:
\bea
\frac{j_{\nu,\rm em}}{n_{\rm H}}&=&\int_{a_{\min}}^{a_{\max}} j_{\nu}(a) \frac{dn}{n_{\H}da} da,\label{eq:jem}\\
\frac{j_{\nu,\rm pol}}{n_{\rm H}}&=&\int_{a_{\min}}^{a_{\max}} Q_{J}(a)\cos^{2}\gamma_{B} j_{\nu}(a) \frac{dn}{n_{\H}da} da,\label{eq:jpol}
\ena
where $j_{\nu}(a)$ is the emissivity by a grain of size $a$, $Q_{J}$ is the degree of external alignment, $\gamma_{B}$ is the angle between $\Bv$ and the plane of the sky, which is assumed to be $\gamma_{B}=0$ in the following calculations. Here $a_{\rm min}=3.5$~\AA~ and $a_{\rm max}=100$~\AA~ are the lower and upper cutoff of the PAH size distribution $dn/da$. We use the advanced spinning dust model in \cite{2011ApJ...741...87H} to calculate $j_{\nu}(a)$. Here we adopt the PAH size distribution $dn/da$ from \cite{2001ApJ...548..296W} with the C abundance contained in PAHs of $b_{C}=6\times 10^{-5}$. 

\subsection{Emission and Polarization Spectrum}
\subsubsection{Unidirectional radiation: RNe}
\begin{figure*}
\centering
\includegraphics[width=0.9\textwidth]{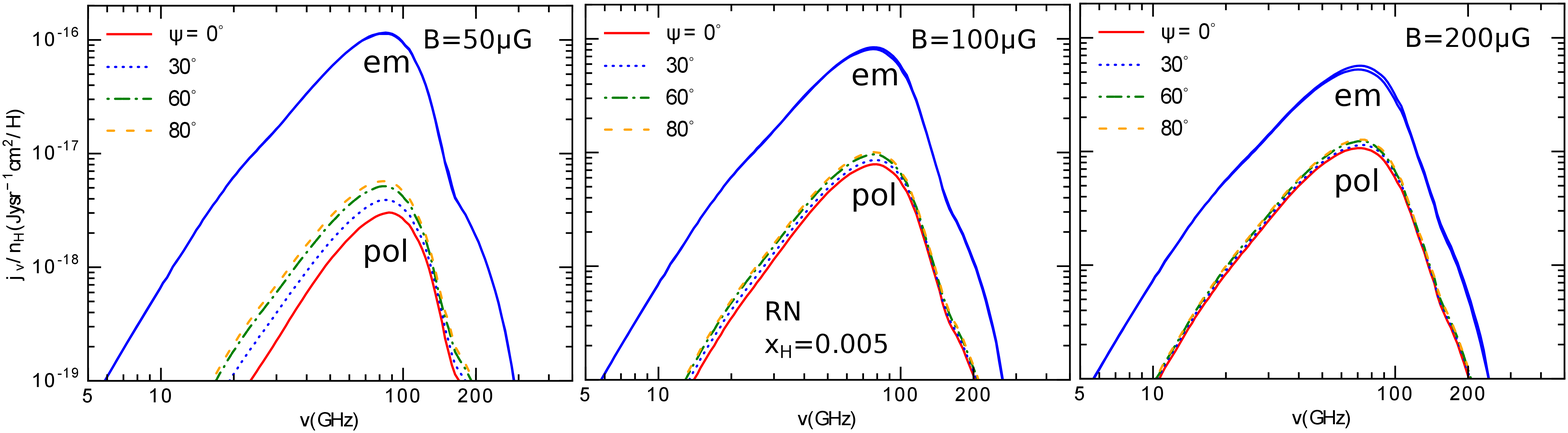}
\includegraphics[width=0.9\textwidth]{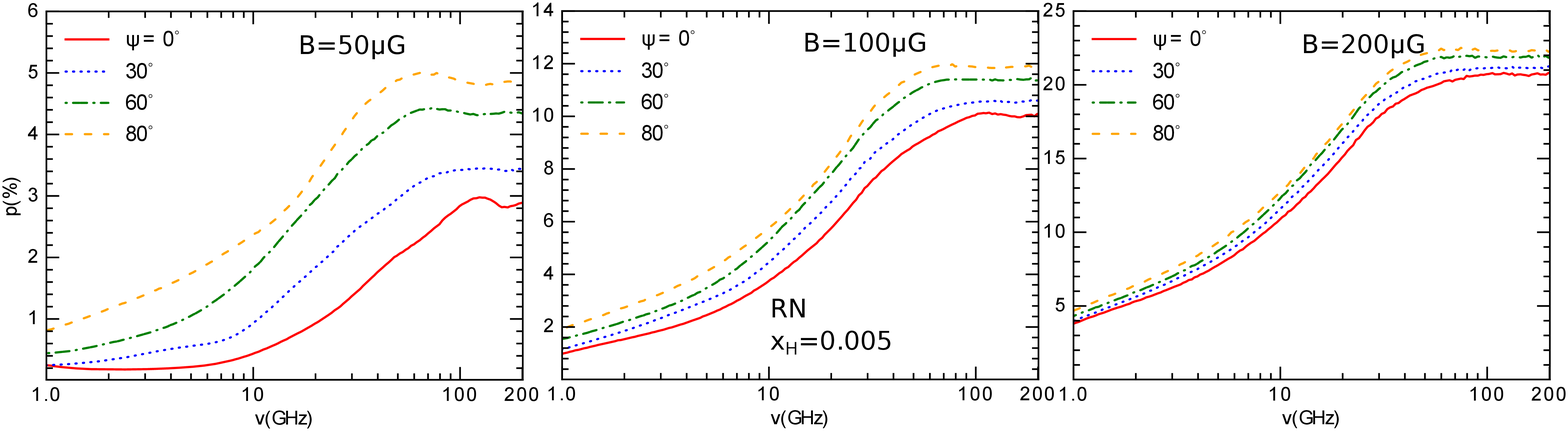}
\caption{Upper panels: unpolarized and polarized spinning dust emissivity by spinning PAHs in RN with $x_{\H}=0.005$ and $U=1000$, computed for the different illumination directions $\psi$. Lower panels: frequency dependence polarization degree of spinning dust emission. Three values of the magnetic field strength, $B=50~\mu$G (left panel), $100~\mu$G (middle panel) and $200~\mu$G (right panel), are considered. }
\label{fig:jnu_PAH}
\end{figure*}

In Figure \ref{fig:jnu_PAH} (upper panels), we present the total spinning emissivity and polarized emissivity from PAHs in the RN, computed for the different illumination angles $\psi$ and ionization fraction $x_{\rm H}=0.005$. The left panel shows the results for $B=50~\mu$G, the middle panel for $B=100~\mu G$ and the right panel for $B=200~\mu$G, respectively.  A slight decrease of the emissivity is found when the field strength is increased, which is a direct consequence of the fast dissipation of rotational energy into heat via magnetic relaxation.

The lower panels show the polarization degree versus the frequency. The polarization increases with the frequency and reaches a nearly saturated value between $p\sim 10-20 \%$ at $\nu> 70$ GHz for $B\ge 100~\mu$G (middle and right panels). The transition frequency of the polarization essentially coincides with the peak frequency of the emissivity, which can facilitate observations. For a given magnetic field, the polarization increases with the illumination angle $\psi$, but the level of such an increase is reduced when the magnetic field is increased. For instance, at $\nu \sim 100$ GHz, the polarization ratio $p(\psi=80^{\circ})/p(\psi=0^{\circ})\sim 1.6$ for $B=50~\mu$G, and it reduces to $p(\psi=80^{\circ})/p(\psi=0^{\circ})\sim 1.05$ for $B=200~\mu$G (see Figure \ref{fig:jnu_PAH}).  

\begin{figure*}
\centering
\includegraphics[width=0.9\textwidth]{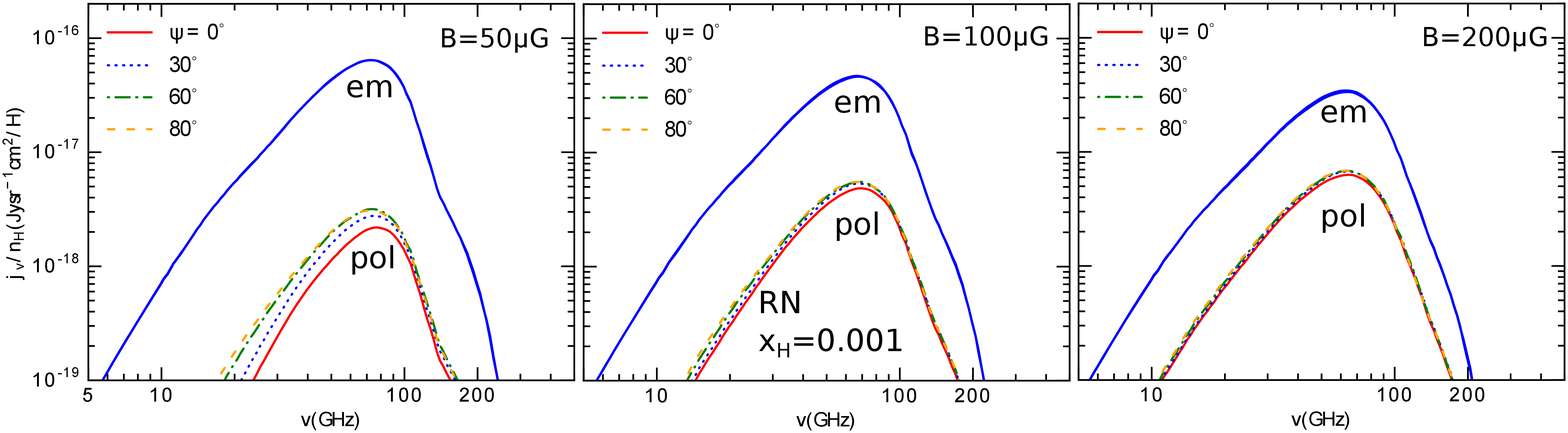}
\includegraphics[width=0.9\textwidth]{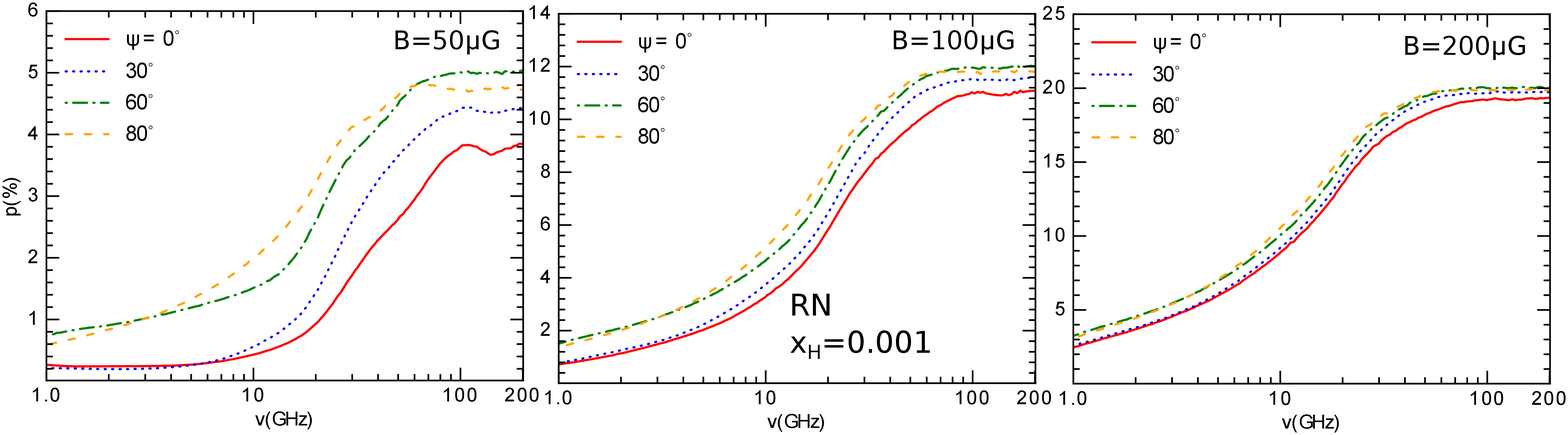}
\caption{Similar to Figure \ref{fig:jnu_PAH}, but for $x_{\rm H}=0.001$.}
\label{fig:jnu_PAH1}
\end{figure*}

\begin{figure*}
\includegraphics[width=0.9\textwidth]{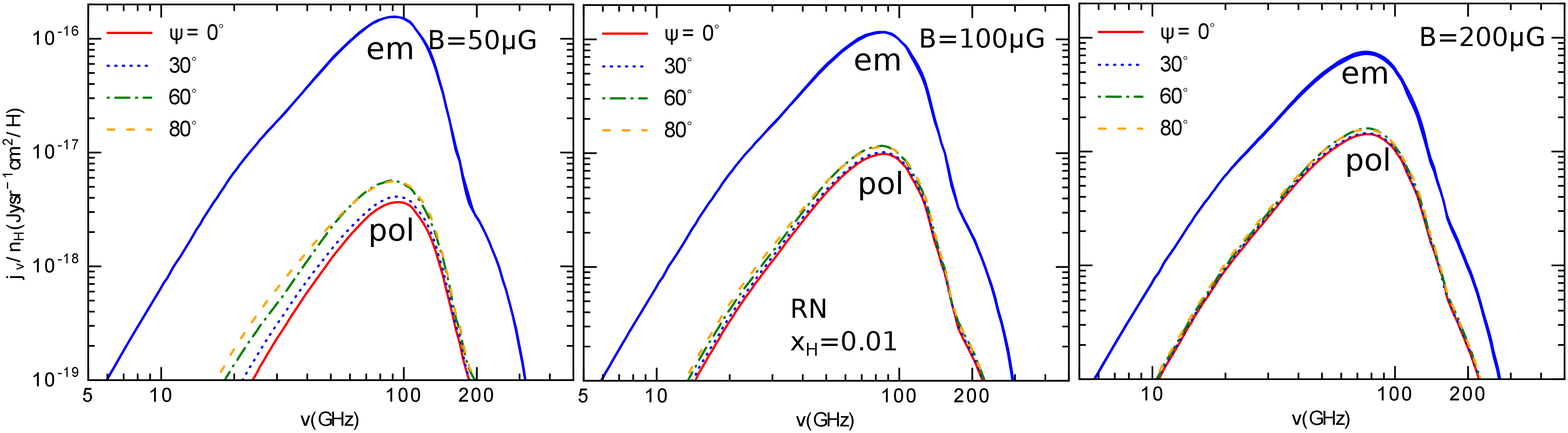}
\includegraphics[width=0.9\textwidth]{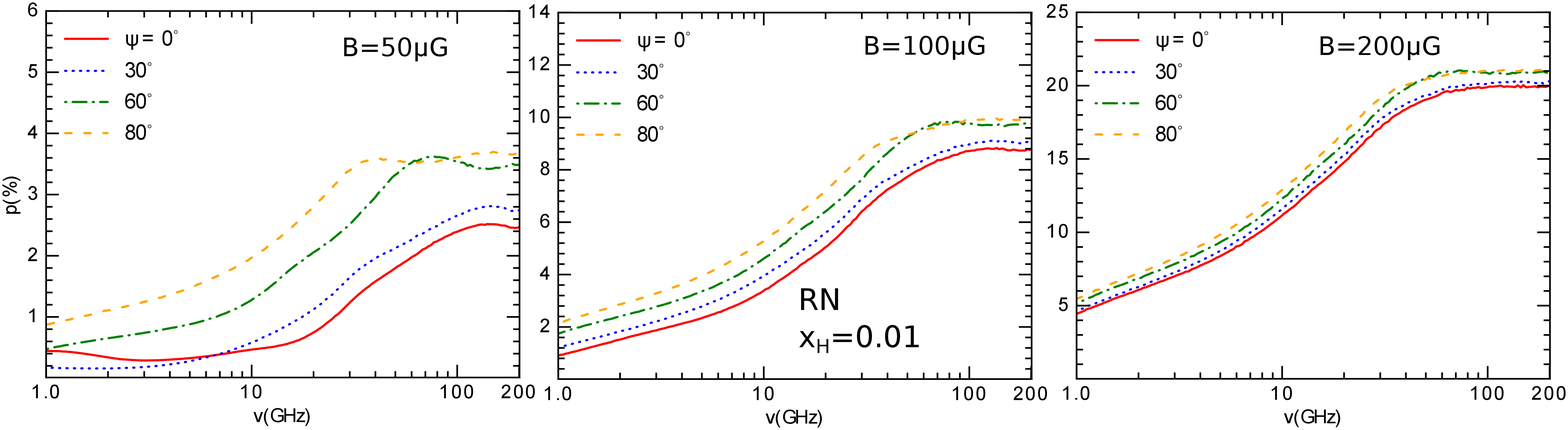}
\caption{Similar to Figure \ref{fig:jnu_PAH}, but for $x_{\rm H}=0.01$.}
\label{fig:jnu_PAH2}
\end{figure*}

Same as Figure \ref{fig:jnu_PAH}, but Figures \ref{fig:jnu_PAH1} and \ref{fig:jnu_PAH2} show the results for two cases, with a lower and higher ionization fractions $x_{\rm H}=0.001$ and $x_{\rm H}=0.01$, respectively. The emissivity is found to vary with $x_{\rm H}$ due to weaker/stronger ion excitation, as expected. The polarization degree increases with $\psi$, in a similar trend as seen in Figure \ref{fig:jnu_PAH} (lower panels).

\begin{figure*}
\centering
\includegraphics[width=0.9\textwidth]{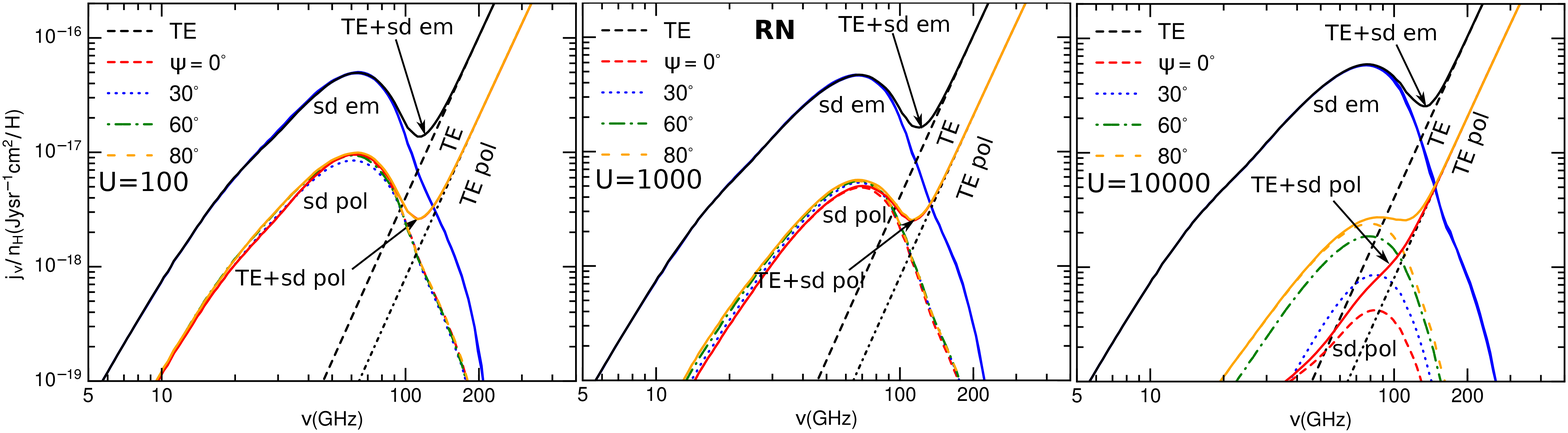}
\includegraphics[width=0.9\textwidth]{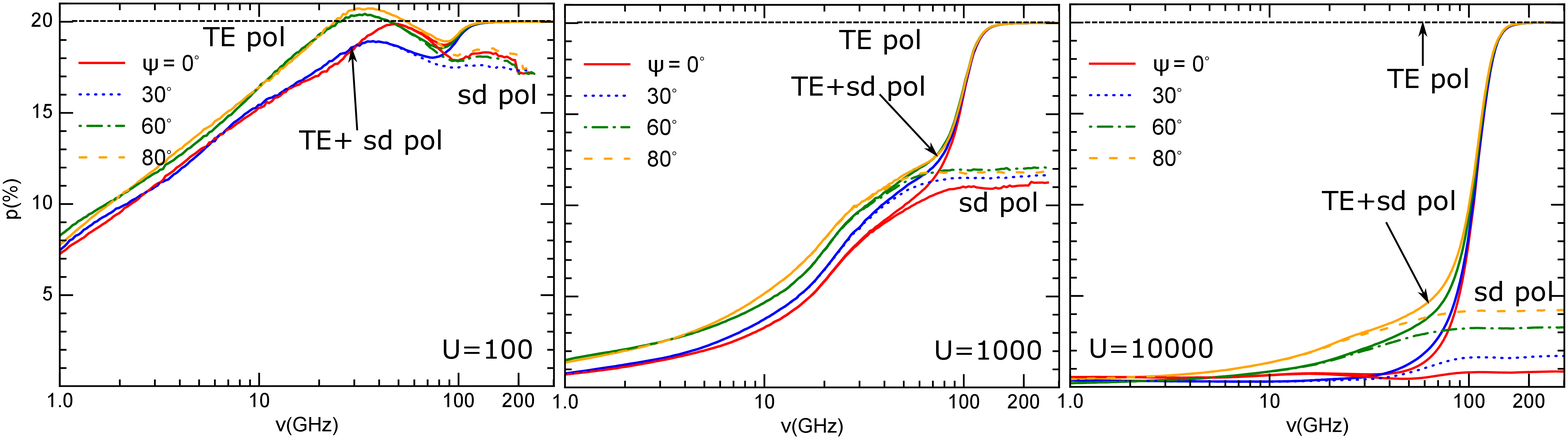}
\caption{Spinning dust emissivity (upper panels) and polarization fraction (lower panels) for RN with different radiation strengths: $U=10^{2}$, $10^{3}$, and $U=10^{4}$. Thermal emission intensity and polarization are also shown in black dashed and dotted lines. The increase of the polarization degree of spinning dust with $\psi$ is more pronounced for $U=10^{4}$.}
\label{fig:jnu_PAH_U}
\end{figure*}

Figure \ref{fig:jnu_PAH_U} (upper panels) shows the total and polarized spinning dust emissivity for three cases of $U=10^{2}, 10^{3}$, and $U=10^{4}$, corresponding to a weak, typical, and a strong radiation field. We also plot the results for thermal dust, both total emission and polarized emission, assuming a fixed polarization fraction of $20~\%$ (see Appendix \ref{apdx:TE}). Spinning dust dominates over thermal dust for $\nu < 150$ GHz, whereas polarized spinning dust dominates at $\nu< 100$ GHz.

Figure \ref{fig:jnu_PAH_U} (lower panels) show the polarization degree from spinning dust, thermal dust, and the total polarization by thermal dust plus spinning dust (solid lines). The polarization degree of spinning dust emission tends to decrease with increasing $U$, which arises from the fact that the stronger radiation induces faster rotational damping by IR emission and thus the lower degree of alignment. However, the increase of the polarization with $\psi$ is much larger, e.g., $p(\psi=80^{\circ})/p(\psi=0^{\circ})\sim 2$ at $\nu \ge 100$ GHz (see the right panel). The total polarization by thermal dust and spinning dust is dominated by thermal dust polarization at $\nu>100$ GHz and by spinning dust at $\nu < 60$ GHz (see solid lines). 

\subsubsection{Unidirectional radiation: PDRs}
\begin{figure*}
\centering
\includegraphics[width=0.9\textwidth]{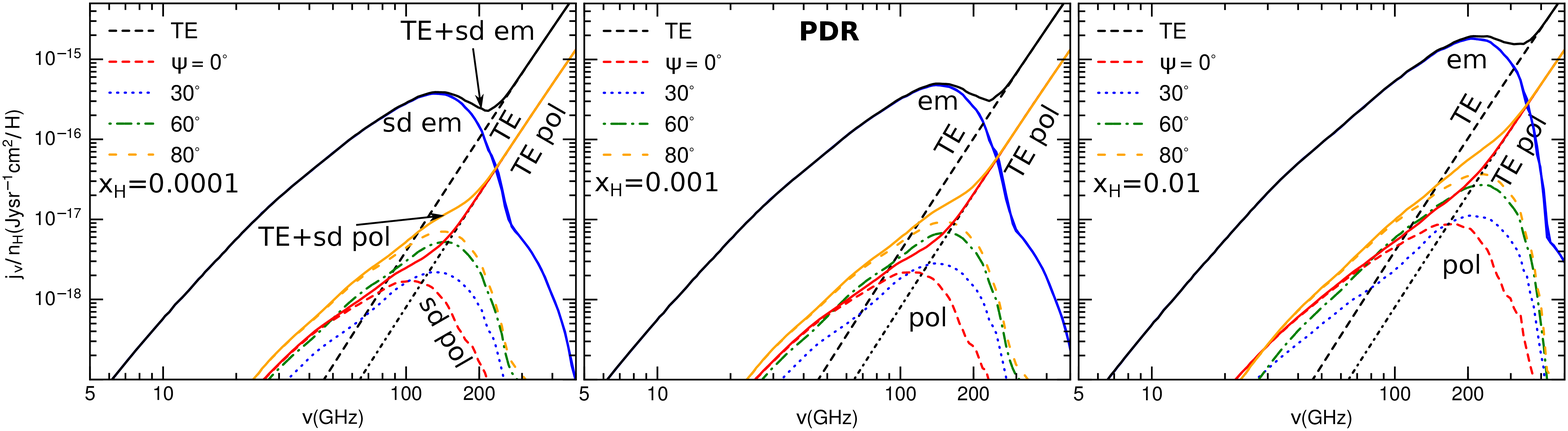}
\includegraphics[width=0.9\textwidth]{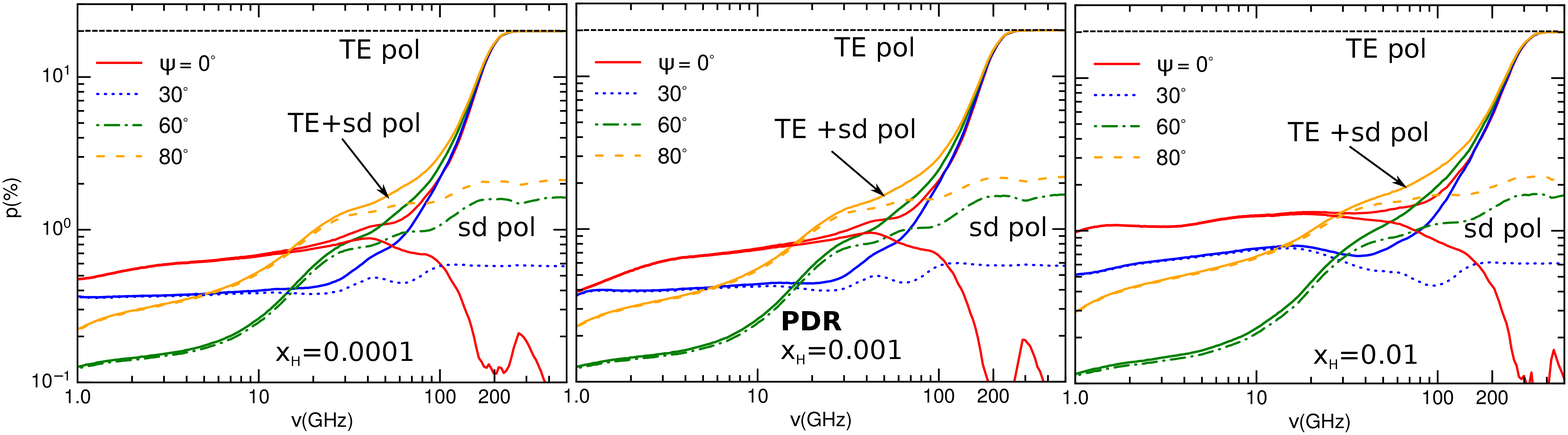}
\caption{Same as Figure \ref{fig:jnu_PAH_U}, but for the standard PDR with varying the ionization fraction $x_{\H}$. Thermal emission intensity and polarization are also shown in black dashed and dotted lines.}
\label{fig:jnu_PAH_PDR}
\end{figure*}

Figure \ref{fig:jnu_PAH_PDR} shows spinning dust emissivity (upper panels) and polarization spectrum (lower panels) computed for the different $\psi$ and $x_{\H}=10^{-4}, 10^{-3}, 10^{-2}$ for the PDR conditions. Thermal dust emission and polarized thermal dust emission is also shown in black dashed and dotted lines, respectively. Spinning dust dominates over thermal dust at frequencies $\nu < 300$ GHz.

As shown in Figure \ref{fig:jnu_PAH_PDR} (lower panels), the polarization of spinning dust can only reach $p \sim 2~\%$ for $\nu>20$ GHz. This polarization level is much lower than in RN (see Figures \ref{fig:jnu_PAH}-\ref{fig:jnu_PAH_U}) because the PDR conditions have much higher IR damping for which PAHs are rotating subthermally and less efficiently aligned according to the magnetic relaxation mechanism (\citealt{2017ApJ...838..112H}). Yet the increase in the polarization with $\psi$ is much more pronounced.\footnote{The polarization below $\sim 0.5~\%$ is within the numerical errors of the Langevin simulations, such that the angle dependence is not discussed.}

The influence of polarized spinning dust emission results in the change in the polarization spectrum at $\nu< 200$ GHz. Indeed, the polarization degree is decreasing with decreasing $\nu$ rather than being flat if only thermal dust emission is considered. 

We note that the present results are obtained assuming the existence of small PAHs in PDRs, with $a_{\min}=3.5$~\AA~ and the peak size of PAH mass distribution $a_{p}\sim 6$~\AA. Such small PAHs are expected to be destroyed by photodissociation in the PDR conditions, leading to an increase in $a_{p}$ with the stellar temperature and luminosity (see \citealt{2017ApJ...835..291S}). As a result, spinning dust emissivity in realistic PDRs is likely lower than shown in Figure \ref{fig:jnu_PAH_PDR} (see \citealt{2011ApJ...741...87H}).

\subsubsection{The diffuse cold neutral medium}
\begin{figure*}
\centering
\includegraphics[width=0.45\textwidth]{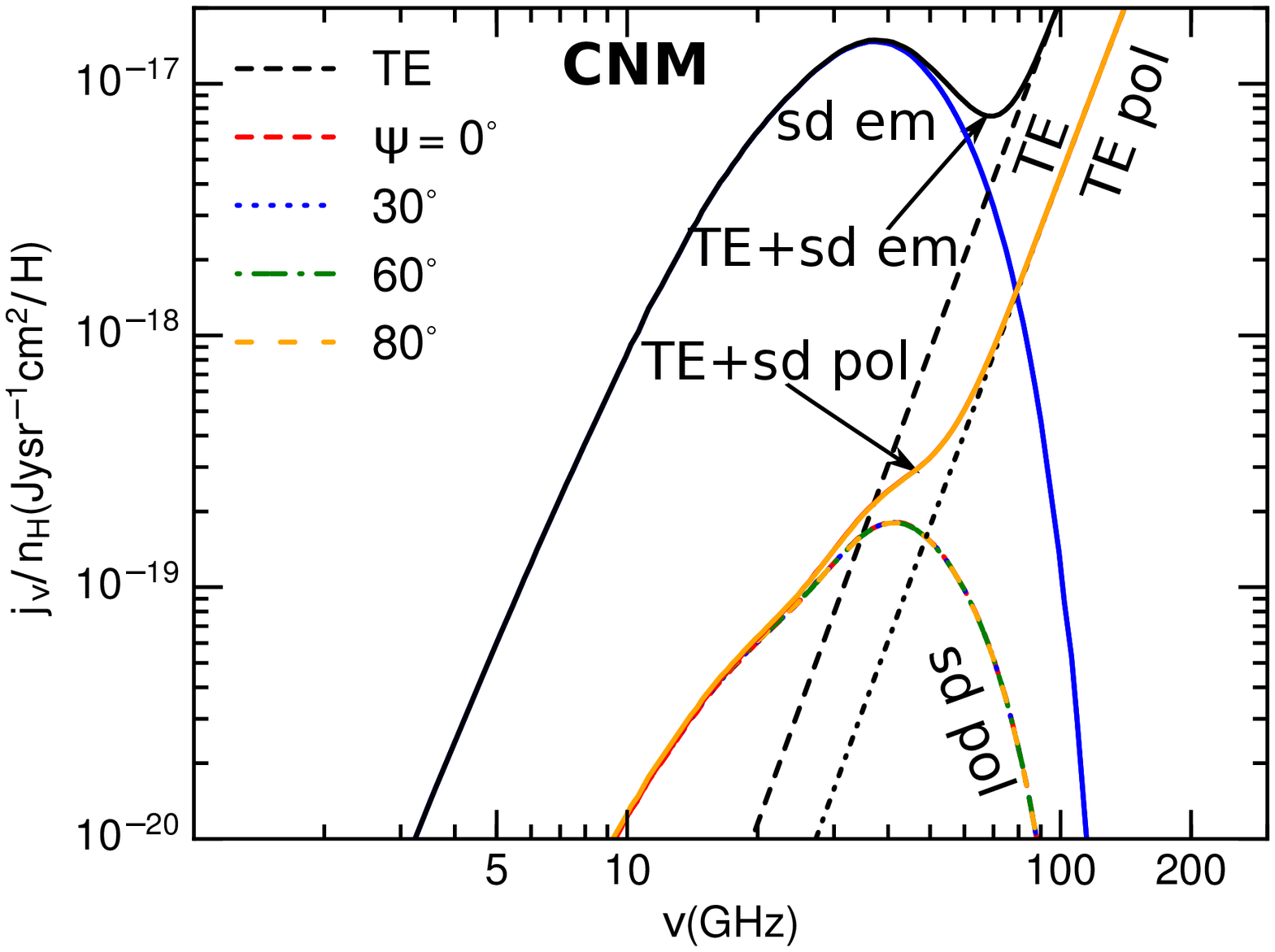}
\includegraphics[width=0.43\textwidth]{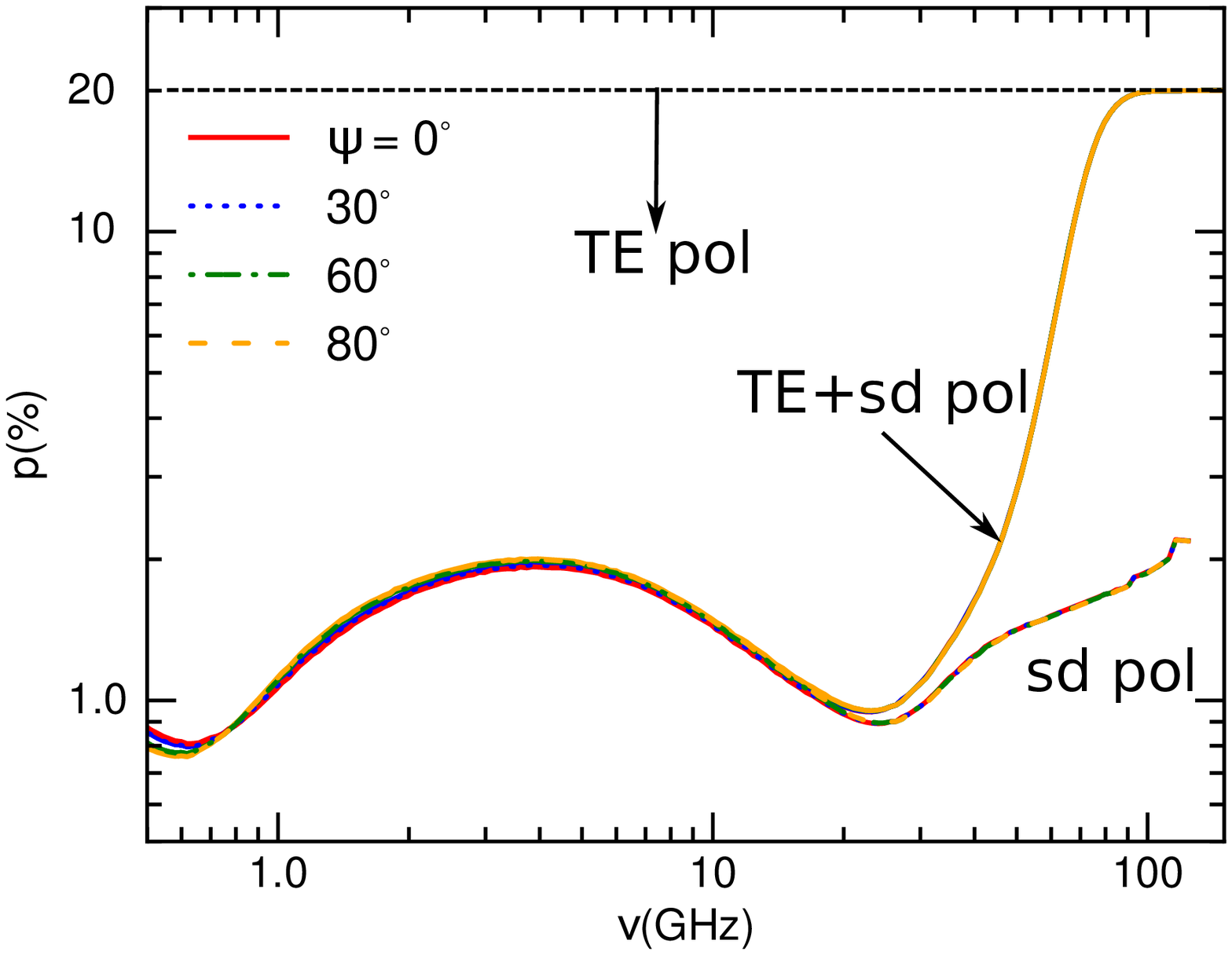}
\caption{Spinning dust emissivity (left panel) and polarization spectrum (right panel) for the CNM with a degree of anisotropy of $\gamma=0.1$.}
\label{fig:jnu_PAH_CNM}
\end{figure*}

For sake of completeness, in Figure \ref{fig:jnu_PAH_CNM} we show the results for the CNM with the low anisotropy degree of $\chi=0.1$. Both emissivity and polarization show an negligible increase with increasing $\psi$. The degree of polarization is low, peak at $\nu\sim 3$ GHz and becomes smaller than $1.5~\%$ for $\nu> 20$ GHz.

The spinning dust is dominant over thermal dust at $\nu < 70$ GHz. The polarized spinning dust can slightly increase the total polarized emission from thermal dust. 

\section{Discussion}
\subsection{Alignment of PAHs in an anisotropic radiation field: magnetic relaxation vs. cross-section}
Resonance paramagnetic relaxation was first suggested by \cite{2000ApJ...536L..15L} as a way to align PAHs and paramagnetic nanoparticles. Later, numerical calculations in \cite{2014ApJ...790....6H} found that resonance relaxation can enable PAHs to be aligned up to a degree of $\sim 10~\%$ (see also \citealt{2017ApJ...838..112H}). These studies assumed that PAHs are illuminated by the isotropic diffuse interstellar radiation and disregard the effect of anisotropic radiation field. 

\cite{Lazarian:1995p3034} suggested a cross-sectional mechanism for aligning classical grains subject to a gas flow. The essence of the mechanism is that the rate of the diffusion of the grain angular momentum can change depending on the cross-section that the planar grain exposes to the flow of atoms, resulting in the alignment of grains that minimizes its cross section. The cross-sectional alignment for classical grains requires the directed flow of hydrogen atoms, which may be difficult to get for ultrasmall grains which usually move with relatively small velocities in respect to the gas. However, the differences in cross sections (see Section \ref{sec:FGIR}) are important for the radiation that interacts with PAHs. This induces a similar effect that the grain stays statistically longer in the position of its minimal cross section.\footnote{Other mechanical processes of alignment e.g. Gold alignment (\citealt{Gold:1952p5848}; \citealt{1976Ap&SS..43..291D}; \citealt{1994MNRAS.268..713L}) also formally make the cross section of the grain interaction with the flow minimal. However, this is done through the depositing of the angular momentum with the grain. The cross-section mechanism is based on different physics. It is based on the dependence of the diffusion of the angular momentum on the grain-flow cross section. In \cite{1995MNRAS.277.1235L} this was due to the changing the rate of cross overs depending on the cross section, in the present paper this is due to the change of the rate of the photon emission depending on grain-radiation cross-section.}

In this paper, we simultaneously treat PAH alignment by paramagnetic relaxation and cross-sectional mechanism in the presence of anisotropic radiation fields. To identify the efficiency of the both mechanisms, we perform calculations for the different magnetic field strengths and illumination directions. We have quantified the alignment of PAHs in three environment conditions, including RN, PDR, CNM, with various physical parameters. For the unidirectional radiation fields (RN and PDR with anisotropy degree $\gamma=1$), we demonstrate that the anisotropic radiation can increase the degree of PAH alignment, with the alignment degree increasing with the angle between the radiation direction and the magnetic field. This is expected from the cross-sectional mechanism acting on the photon-grain interaction. The cross-sectional process is more efficient when the magnetic field is lower, thus one may estimate the efficiency of cross-sectional mechanism based on the alignment efficiency obtained for a very weak magnetic field. In this sense, we found that the alignment efficiency $Q_{J}$ cannot exceed $5~\%$ for $B=50~\mu$G in the RN conditions (see the right panels in Figures \ref{fig:RQJ_RN1}-\ref{fig:RQJ_RN2}).

Moreover, among the considered environments, we found that PAHs are more efficiently aligned in the RN conditions where the gas density is not as high as in PDRs, which originates from the resonance paramagnetic relaxation mechanism. A low degree of PAH alignment in the CNM (i.e., $\gamma=0.1$) obtained in this study is consistent with our previous studies where isotropic radiation field is assumed \citep{2014ApJ...790....6H}. 

\subsection{Polarization of spinning PAH emission and comparison with observations}
Using the obtained degree of PAH alignment, we calculate spinning dust emissivity, both total emission and polarized emission. For the diffuse interstellar medium (i.e., CNM), we find that the polarization of spinning PAH emission is about $2\%$ peak at $5$GHZ and declines rapidly to below $1\%$ at $\nu > 20$ GHz. 

For the RN conditions, we find that the polarization degree can be large, increasing from $5~\%$ at $10$ GHz to $20~\%$ at $100$ GHz. The peak polarization is close to the peak of emission intensity, which favors the detection of the polarization. The difference arises from the fact that in the RN, the degree of alignment increases with the PAH size for $a<10~\AA$, whereas the alignment decreases rapidly for the CNM (see Figures \ref{fig:RQJ_RN1} and \ref{fig:RQJ_CNM}). Such a difference arises from the excitation by ion collisions that dominates the rotational excitation in the RN for smallest PAHs with negatively charges. The effect of cross-sectional mechanism is particularly important for the RN and PDR conditions. We find that for the RN the polarization degree can be increased by $200~\%$ when the illumination direction is changed from the parallel to nearly perpendicular to the magnetic field. In the PDR, the polarization is lower because of the dominant damping and randomization by gas collisions, but the efficiency of cross-sectional mechanism is stronger (see Section \ref{sec:polem}). 

The theoretical predictions for the polarization of spinning PAH emission in the CNM appear to be consistent with the available observational data. For instance, several observational studies (\citealt{2006ApJ...645L.141B}; \citealt{2009ApJ...697.1187M}; \citealt{2011MNRAS.418L..35D}; \citealt{Battistelli:2015dt}) have found the upper limits for the AME polarization to be between $1~\%$ and $5~\%$. An upper limit of $1~\%$ is also reported in various studies (see \citealt{LopezCaraballo:2011p508}; \citealt{RubinoMartin:2012ji}; \citealt{2017MNRAS.464.4107G}). Latest results by Planck (\citealt{2015arXiv150606660P}) show the AME polarization to be $0.6\pm 0.5~\%$, consistent with our calculations for the CNM. Observations for the polarization of AME from RNe are not yet available. Nevertheless, in the light of this study, we suggest that future observations for the polarization of AME should search for polarized spinning dust emission in the RN conditions. Note that our theoretical results are obtained assuming the magnetic field perpendicular to the line of sight. It is known that the interstellar medium is magnetized and turbulent. As a result, the magnetic field wandering can reduce the polarization degree presented in this paper.

\subsection{Implications for polarized infrared emission and tracing magnetic fields}

If PAHs are indeed aligned with the magnetic fields, then, we can probe the magnetic fields in the RN conditions, such as circumstellar disks around T-Tauri and Herbig Ae/Be stars via mid-IR polarization (\citealt{2017ApJ...838..112H}). Moreover, expecting the same alignment of nanosilicates by resonance relaxation, the polarization of $\sim 10 \mum$ silicate emission feature is perhaps comparable to that of the 11.3 $\mum$ feature. RAT alignment theory predicts that very large grains are not aligned with respect to the magnetic field because the Larmor precession becomes slower than gas randomization (see Eq. 41 in \citealt{2016ApJ...831..159H}). Instead, these very large grains are predicted to be aligned with respect to the anisotropic radiation direction \citep{2007MNRAS.378..910L}. The incorporation of iron inclusions in dust grains can enhance the maximum size of grains which are still be coupled to the magnetic field (\citealt{2016ApJ...831..159H}). Such a maximum size as a function of the local gas density, magnetic field strength, and iron inclusions can be estimated using Eq. 42 in \citealt{2016ApJ...831..159H}. We note that the alignment along the radiation direction is first theoretically studied by \cite{2007MNRAS.378..910L} and is modeled by \cite{2017ApJ...839...56T} for protoplanetary disks. As a result, mm wavelength polarization may ease to trace the magnetic field in circumstellar disks. ALMA observations of HL Tau at 3.1 mm by \cite{Kataoka:2017fq} and at 1.3 mm by \cite{Stephens:2017ik} show that the polarization patterns are azimuthal, as expected from the polarization produced by grains aligned with the radiation direction (\citealt{2017ApJ...839...56T}). Thus, mid-IR polarization appears to be a useful way to trace magnetic fields in the surface layer of circumstellar disks where grains are expected to be aligned with the magnetic field (see e.g., \citealt{2016ApJ...832...18L}).

\subsection{Effect of temperature fluctuations on nanoparticle alignment in intense radiation environments}\label{sec:Tfluc}

In the present paper, to capture the effect of anisotropic illumination on PAH alignment, the temperature of PAHs is assumed to be fixed. In realistic conditions, the temperature of small PAHs undergo strong fluctuations, with frequent thermal spikes after single-photon absorption (\citealt{1989ApJ...345..230G}). Let us now discuss the effect of temperature fluctuations on PAH alignment.

For a typical PAH of size 10~\AA, the mean time between two UV photon absorption events is $\tau_{\rm abs}\sim (10^{6}/U)$ s, while the radiative cooling time via IR emission is $\tau_{\rm rad}\sim 100$ s (\citealt{2001ApJ...551..807D}). In the RN and PDR conditions with $U \gtrsim 10^{3}$, one has $\tau_{\rm rad}\gtrsim \tau_{\rm abs}$, for which the PAH molecule cannot cool completely before a subsequent UV photon absorption. Therefore, the PAH temperature undergoes frequent fluctuations, but the amplitude of fluctuations is smaller than in the case of the standard diffuse radiation field with $\tau_{\rm rad}\ll \tau_{\rm abs}$ (i.e., CNM). For PAHs smaller than 10~\AA, the effect of thermal spikes is more important because the radiation time becomes much shorter than the absorption time (see \citealt{2001ApJ...551..807D}).

The magnetic alignment timescale by paramagnetic relaxation is $\tau_{\rm mag} \sim 0.5 (a/10\AA)^{2}(B/100\mu \rm G)^{-2}(1.2\times 10^{-13}\s/K(\omega))$ yr where $K(\omega)=\chi_{2}/\omega$ with $\chi_{2}$ the imaginary part of magnetic susceptibility (see \citealt{2016ApJ...831..159H}). This timescale is much longer than the thermal spike timescale as well as internal relaxation. Therefore, an adequate treatment of grain alignment needs to average the damping and diffusion coefficients over the grain temperature distribution (e.g., \citealt{2016ApJ...831...59D}). Such a study is beyond the scope of this paper, which is aimed to quantify the effect of cross-sectional alignment only. Nevertheless, we expect temperature fluctuations to reduce the degree of internal alignment and external alignment because of stronger thermal fluctuations during the spikes. For instance, \cite{2017ApJ...838..112H} show a significant decrease in the alignment degree with increasing the grain temperature prior UV photon absorption $T_{0}$.

\section{Summary}\label{sec:concl}

We have quantified the cross-sectional mechanism of grain alignment that is based on the difference in the rate of the photon absorption by PAHs depending on its orientation relative to the photon flux and computed the polarization of microwave emission arising from rapidly spinning PAHs. Our principal results are summarized as follows:

\begin{itemize}

\item[1] We calculated the averaged diffusion coefficients by IR emission over the torque-free motion of PAHs, thermal fluctuations, and fast Larmor precession. We demonstrated that the averaged coefficients exhibit anisotropy for suprathermal rotation, but nearly isotropic for subthermal rotation. 

\item[2] We numerically computed the degree of PAH alignment by varying the illumination angle $\psi$ relative to the magnetic field. For the unidirectional radiation fields of RN and PDR, we find that the degree of alignment tends to increase with increasing $\psi$. For the diffuse ISM (CNM), the variation of alignment with $\psi$ is negligible due to a low degree of anisotropy. The alignment of PAHs in the RN conditions is the most efficient compared to that in the CNM and PDR.

\item[3] Our numerical results demonstrate that the cross-sectional alignment can also work for photon-grain interactions in an anisotropic radiation field, but the resulting alignment efficiency is moderate. The resonance paramagnetic relaxation still plays a dominant role in aligning PAHs with the magnetic field. The obtained degree of PAH alignment can successfully explain the polarization of 2$\%$ observed at 11.3$~\mum$ from the MWC 1080 nebula by \cite{Zhang:2017ea}, as shown in \cite{2017ApJ...838..112H}.

\item[4] We predicted the polarization degree of spinning PAH emission in the RN, PDR, and CNM. We found that the polarization degree is low for the diffuse regions (CNM) and PDR, but can be large, up to $20\%$, for the RN conditions and tends to increases with the illumination direction. The peak polarization is close to the peak emission, which reveals that RN is the most favored conditions for detecting the spinning dust polarization.

\end{itemize}

\acknowledgments
We thank anonymous referees for useful comments that help improve our paper. T.H. acknowledges the support by the Basic Science Research Program through the National Research Foundation of Korea (NRF), funded by the Ministry of Education (2017R1D1A1B03035359). 
\appendix

\section{Coordinate systems and transformation}\label{apdx:transform}
\subsection{Definitions of coordinate systems}
Assuming that a PAH is illuminated by radiation from a nearby star with the propagation vector $\hat{k}$.

To account for an external source such as a star, we introduce an frame of reference $\xhat_{k}\yhat_{k}\zhat_{k}$ where $\zhat_{k}$ is directed along the radiation direction $\bk$. 

Let $\ahat_{1},\ahat_{2},\ahat_{3}$ be the grain principal axes where $\ahat_{1}$ is the axis of maximum moment of inertia. For the axisymmetric shape of PAH, $\ahat_{2}$ is chosen along the line of node (see Figure \ref{fig:RFs}).

Let $\xhat_{J}\yhat_{J}\zhat_{J}$ are unit vectors in which $\zhat_{J}$ is parallel to $\bJ$, $\xhat_{J}$ is directed along $\hat{\xi}$, and $\yhat_{J}$ is directed along $\hat{\phi}$ (see Figure \ref{fig:RFs}).  Let $\xhat_{B}\yhat_{B}\zhat_{B}$ be unit vectors defined by the magnetic field in which $\zhat_{B}\| \Bv$.  The orientation of $\bJ$ in the B-frame is determined by two angles $\xi$ and $\phi$.

\subsection{Transformation of coordinate systems}

The alignment of $\bJ$ with respect to the magnetic field is determined by the angles $\xi,\phi$ in the magnetic field coordinate system. The transformation between vector basis $\ehat$ to a new basis $\ehat'$ are done through matrix transformation:
\bea
\ehat_{i} = T_{ij}\ehat_{j},
\ena
where $T$ is the transformation matrix from $\ehat$ to $\ehat'$ reference frame.

For an axisymmetric grain with the angles defined in Figure \ref{fig:RFs}, we have the following (see Appendix A in \citealt{2011ApJ...741...87H}) :
\bea
\ahat_{1}=\cos\theta \zhat_{J} + \sin\theta(\sin\zeta \xhat_{J}-\cos\zeta \yhat_{J}),\label{eq:a1_J} \\
\ahat_{2}=\cos\zeta \xhat_{J}+\sin\zeta \yhat_{J},\\
\ahat_{3}=[\ahat_{1}\times \ahat_{2}]=\sin\theta \zhat_{J} + \cos\theta(-\sin\zeta \xhat_{J}+\cos\zeta \yhat_{J}),\label{eq:a2_J}\\
,\label{eq:a3_J}
\ena
where $\zeta$ denotes the rotation angle around $\ahat_{1}$, e.g., the line of node $N$ can be chosen to be along $\ahat_{2}$.

From $\bJ$-frame to $\Bv$-frame, we get
\bea
\zhat_{J}=\cos\xi \zhat_{B} + \sin\xi(\cos\phi \xhat_{B}+\sin\phi \yhat_{B}),\label{eq:zJ_B}\\
\xhat_{J}=-\sin\xi \zhat_{B} + \cos\xi(\cos\phi \xhat_{B}+\sin\phi \yhat_{B}),\label{eq:xJ_B}\\
\yhat_{J}=[\zhat_{J}\times \xhat_{J}]=\cos\phi \yhat_{B}-\sin\phi \xhat_{B},\label{eq:yJ_B}
\ena
where $\yhat_{J}$ lies in the $\xhat_{B}\yhat_{B}$ plane.

The direction of $\Bv$ in k-frame is determined by the angle $\psi$ and $\zeta$ to consider for 3D magnetic field. The transformation from B-frame to k-frame is given by
\bea
\zhat_{B}=\cos\psi \zhat_{k} + \sin\psi(\cos\zeta \xhat_{k}+\sin\zeta \yhat_{k}),\label{eq:zB_k}\\
\xhat_{B}=-\sin\psi \zhat_{k} + \cos\psi(\cos\zeta \xhat_{k}+\sin\zeta \yhat_{k}),\label{eq:xB_k}\\
\yhat_{B}=[\zhat_{B}\times \xhat_{B}]=\cos\zeta \yhat_{k}-\sin\zeta \xhat_{k}.\label{eq:yB_k}
\ena

It follows that 
\bea
\cos\beta=\zhat_{J}.\zhat_{k}=\cos\xi\cos\psi -\sin\xi\cos\phi\sin\psi,\\
\sin\beta\cos\varphi =\zhat_{J}.\xhat_{k}=\cos\xi\sin\psi\cos\zeta + \sin\xi\left(\cos\phi\cos\psi\cos\zeta -\sin\phi\sin\zeta \right)
\ena

Let $\psi$ be the angle between $\zhat_{k}$ and $\zhat_{B}$. Then, the angle $\Theta$ between $\ahat_{1}$ and $k$ ($\zhat_{k}$) is then given by
\bea
\cos\Theta = \ahat_{1}.\zhat_{k} =\cos\theta \zhat_{J}.\zhat_{k} +\sin\theta (\sin\zeta \xhat_{J}.\zhat_{k}-\cos\zeta \yhat_{J}.\zhat_{k}).\label{eq:a1_zk}
\ena
From Equations (\ref{eq:zJ_B})- (\ref{eq:yB_k}), we obtain
\bea
\zhat_{J}.\zhat_{k} = \cos\xi\cos\phi - \sin\xi\cos\phi\sin\psi,\\
\xhat_{J}.\zhat_{k} = -\sin\xi\cos\phi - \cos\xi\cos\phi\sin\psi,\\
\yhat_{J}.\zhat_{k} = \sin\phi\sin\psi.\label{eq:yj_zk}
\ena

\subsection{Transformation of the diffusion coefficients}
Note that the final diffusion coefficients in the lab system are already presented in \cite{Hoang:2010jy}. Here we present the derivation of these coefficients by following the transformation of the damping and diffusion coefficients from the grain body frame to the lab frame defined by the magnetic field. 

\subsubsection{Body system to $\bJ$-frame and averaging over the grain precession}
Let consider the transformation of the diffusion coefficients from the body frame to the $\bJ$-frame. In the body frame, the damping coefficients for an axisymmetric grain (oblate) are described 
\bea
A_{z}^{b}\equiv \langle\frac{\Delta J^{b}_{z}}{\Delta t}\rangle =  -\frac{J_{z}^{b}}{\tau_{1}},\label{eq:dJzdt_b}\\
A_{x,y}^{b}\equiv \langle\frac{\Delta J^{b}_{x,y}}{\Delta t}\rangle =  -\frac{J_{x,y}^{b}}{\tau_{2}},\label{eq:dJxydt_b}
\ena
where $\tau_{1}=\tau_{\|},\tau_{2}=\tau_{\perp}$ denote the characteristic damping timescale for the rotation along the grain symmetry and perpendicular axis, respectively. 

The increment of $\bJ$ after a time interval $\Delta t$ is evaluated in the body frame as $\Delta {\bJ}^{b}$. For the body frame defined with $\ahat_{2}$ along the line of node, the vector $\Delta {\bJ}^{b}$ can be transformed to the $J$-frame as follows:
\bea
\Delta J_{z} = \Delta J_{z}^{b}\cos\theta +\Delta J_{y}^{b}\sin\theta,\\
\Delta J_{x} = \Delta J_{x}^{b}\cos\phi -\Delta J_{y}^{b}\cos\theta\sin\phi + \Delta J_{z}^{b}\sin\theta\sin\phi,\\
\Delta J_{y} = \Delta J_{x}^{b}\sin\phi +\Delta J_{y}^{b}\cos\theta\cos\phi - \Delta J_{z}^{b}\sin\theta\sin\phi.
\ena

Averaging over the precession angle $\phi$ of the grain around $\ahat_{1}$, we obtain
\bea
A_{x,y}^{J}\equiv \langle\frac{\Delta J_{x,y}}{\Delta t}\rangle = 0, \\
A_{z}^{J}\equiv \langle\frac{\Delta J_{z}}{\Delta t}\rangle= \langle \frac{\Delta J_{z}^{b}}{\Delta t}\rangle\cos\theta + \langle\frac{\Delta J_{z}^{b}}{\Delta t}\rangle\sin\theta=-\frac{J_{z}^{b}}{\tau_{1}}\cos\theta - \frac{J_{y}^{b}}{\tau_{2}}\sin\theta,
\ena

With the choice of $\bJ$ so that the $\bJ$ is directed along $\zhat$, we have $\bJ  = J_{z}^{b}\hat{z}_{b} + J_{y}^{b}\hat{y}_{b}$, $J_{z}^{b}=J_{z}\cos\theta$, and $\tan\theta = J_{y}^{b}/J_{z}^{b}$. Therefore, we obtain
\bea
\langle \frac{\Delta J_{z}}{\Delta t}\rangle &=&  -\frac{J_{z}^{b}}{\tau_{1}}\left(\cos\theta + \frac{\tau_{1}}{\tau_{2}}\sin\theta\tan\theta\right)=\langle \frac{\Delta J_{z}^{b}}{\Delta t}\rangle\left(\cos\theta + \frac{\tau_{1}}{\tau_{2}}\sin\theta\tan\theta\right)\\
&=&-\frac{J_{z}^{b}}{\cos\theta\tau_{1}}\left(\cos^{2}\theta + \frac{\tau_{1}}{\tau_{2}}\sin^{2}\theta\right)=-\frac{J_{z}}{\tau_{1}}\left(\cos^{2}\theta + \frac{\tau_{1}}{\tau_{2}}\sin^{2}\theta\right)=-\frac{J_{z}}{\tau_{\rm eff}},\label{eq:dJdz_aJ}
\ena
where 
\bea
\tau_{\rm eff} =\tau_{1}\epsilon(\theta),~\epsilon(\theta)= \frac{1}{\cos^{2}\theta + \gamma\sin^{2}\theta}, ~ \gamma=\frac{\tau_{1}}{\tau_{2}}.\label{eq:taueff}
\ena

The above equation corresponds to the transformation of the damping coefficient:
\bea
A_{z}^{J} = A_{z}^{b}\frac{\left(\cos^{2}\theta + \frac{\tau_{1}}{\tau_{2}}\sin^{2}\theta\right)}{\cos\theta}.
\ena

Similarly, after averaging the diffusion coefficients over the grain precession, we obtain the diffusion coefficients in the $\bJ$-system.
\bea
B_{xx}^{J}\equiv \langle\frac{(\Delta J_{xx})^{2}}{\Delta t}\rangle = \frac{1}{2}\left[\langle \frac{(\Delta J_{x}^{b})^{2}}{\Delta t}\rangle (1+\cos^{2}\theta) + \langle \frac{(\Delta J_{z}^{b})^{2}}{\Delta t}\rangle \sin^{2}\theta \right],\label{eq:Bxx_J}\\
B_{zz}^{J}\equiv\langle \frac{(\Delta J_{z})^{2}}{\Delta t}\rangle =\left[\langle \frac{(\Delta J_{z}^{b})^{2}}{\Delta t}\rangle \cos^{2}\theta + \langle \frac{(\Delta J_{y}^{b})^{2}}{\Delta t}\rangle \sin^{2}\theta \right],\label{eq:Bzz_J}
\ena
and $B_{yy}^{J}=B_{xx}^{J}$.

Equations (\ref{eq:dJdz_aJ}-\ref{eq:dJJdz_aJ}) can be used to find the diffusion coefficients in $\bJ$-frame for an arbitrary interaction process. For gas-grain collisions, we get $\tau_{1}=\tau_{\gas,1}$ and $\tau_{2}=\tau_{\gas,2}$, which are $\tau_{\H,\|}$ and $\tau_{\H,\perp}$ for a purely hydrogen atomic gas.

For IR emission, let recall that $\Delta J_{\rm IR,i}^{b}/\Delta t = -J_{i}^{b}/\tau_{\rm IR,i}$. Then, $\tau_{1}=\tau_{\rm IR,1}$ and $\tau_{2}=\tau_{\rm IR,2}$ where $\tau_{\rm IR,i}^{-1}=F_{\rm IR,i}\tau_{gas,i}^{-1}$. Thus, we can get the IR damping in the $J$-frame
\bea
\langle \frac{\Delta J_{\rm IR,z}}{\Delta t}\rangle = -\frac{J_{z}}{\tau_{\rm IR,eff}}=-\frac{J_{z}}{\tau_{\rm eff}}\times F_{\rm IR,1},
\ena
where
\bea
\tau_{\rm IR,eff} =  F_{\rm IR,1}^{-1}\tau_{\gas,1}\epsilon_{\rm IR}(\theta),~\epsilon_{\rm IR}(\theta)= \frac{1}{\cos^{2}\theta + \gamma_{\rm IR}\sin^{2}\theta}, ~ \gamma_{\rm IR}=\frac{\tau_{\rm IR,1}}{\tau_{\rm IR,2}}= \frac{\tau_{\gas,1}}{\tau_{\gas,2}} \times\frac{F_{\rm IR,2}}{F_{\rm IR,1}}      
\ena

For a simultaneous transformation of all damping processes with individual timescales $\tau_{i}$. Then, we have the total rate of damping: $\tau_{\rm tot,1}^{-1}=\sum_{i=n,ion,p,IR} \tau_{i,1}^{-1}=(\sum_{i}F_{i,1})\tau_{\gas,1}^{-1}=F_{\rm tot,1}\tau_{\gas,1}^{-1}$, and $\tau_{\rm tot,2}^{-1}=\sum_{i=n,ion,p,IR} \tau_{i,2}^{-1}=(\sum_{i}F_{i,2}) \tau_{\gas,2}^{-1}=F_{\rm tot,2}\tau_{\gas,2}^{-1}$. Thus, the total damping rate in $J$-system is
\bea
A_{z}^{J} \equiv \langle\frac{\Delta J_{z}}{\Delta t}\rangle = -\frac{J_{z}}{\tau_{\rm tot,eff}}, ~ A_{x,y}^{J}=\langle\frac{\Delta J_{x,y}}{\Delta t}\rangle=0,
\ena
where $\tau_{\rm tot,eff}=\tau_{\rm tot,1}\epsilon_{\rm tot}$. Following Equation (\ref{eq:taueff}), it yields
\bea
\tau_{\rm tot,eff}=\tau_{\gas,1}F_{\rm tot,1}^{-1}\epsilon_{\rm tot},~\epsilon_{\rm tot}(\theta)= \frac{1}{\cos^{2}\theta + \gamma_{\rm tot}\sin^{2}\theta}, ~ \gamma_{\rm tot}=\frac{\tau_{\rm tot,1}}{\tau_{\rm tot,2}}= \frac{\tau_{\gas,1}}{\tau_{\gas,2}} \times\frac{F_{\rm tot,2}}{F_{tot,1}}.      
\ena

Let $B_{tot,ii}^{b}=\sum_{j=n,p,ion,IR} B_{j,ii}^{b}$ be the total diffusion coefficients from the various processes in the body frame. The total diffusion coefficients in the $\bJ$-frame then can be given by Equation (\ref{eq:dJJdz_aJ}), but with
\bea
B_{xx}^{J}\equiv \langle\frac{(\Delta J_{x,y})^{2}}{\Delta t}\rangle = \frac{1}{2}\left[B_{tot,xx}^{b}(1+\cos^{2}\theta) + B_{tot,zz}^{b} \sin^{2}\theta \right],\\
B_{zz}^{J}\equiv \langle \frac{(\Delta J_{z})^{2}}{\Delta t}\rangle =\left[ B_{tot,zz}^{b} \cos^{2}\theta + B_{tot,yy}^{b}\sin^{2}\theta \right],\label{eq:dJJdz_aJ}
\ena
and $B_{yy}^{J}=B_{xx}^{J}$. Note that the diffusion coefficients are proportional to the excitation coefficients:
\bea
B_{zz}^{b}= G_{zz}\left(\frac{2I_{1}kT_{\gas}}{\tau_{\gas,1}} \right), B_{xx}^{b}= G_{xx}\left(\frac{2I_{2}kT_{\gas}}{\tau_{\gas,2}} \right),
\ena
where $G_{xx}$ and $G_{zz}$ are calculated in the grain body system. The diffusion and excitation coefficients are defined by energy, so they are additive. In dimensionless units, we have (see also \citealt{2016ApJ...821...91H})
\bea
B_{zz}^{'b}= B_{zz}^{b}\left(\frac{\tau_{\gas,1}}{2I_{1}kT_{\gas}} \right)=G_{zz}, ~B_{xx}^{'b}= B_{xx}^{b}\left(\frac{\tau_{\gas,1}}{2I_{1}kT_{\gas}} \right)=G_{xx}\left(\frac{\tau_{\gas,1}}{h_{a}\tau_{\gas,2}} \right),
\ena

\subsubsection{$\bJ$-frame to $\Bv$-frame and Averaging over the Larmor precession}
Since $A_{x,y}=0$, following Equation (\ref{eq:dJJdz_aJ}), the damping coefficient is directly given by 
\bea
A_{x,y,z} = -\frac{J_{x,y,z}}{\tau_{\rm tot,eff}}.
\ena

The transformation of the diffusion coefficients from $\bJ$-frame to $\Bv$-frame is equivalent to the transformation from body-frame to $\bJ$-frame, where $\theta,\zeta$ in Equations (\ref{eq:Bxx_J}) and (\ref{eq:Bzz_J}) will be replaced by $\xi$ and $\phi$. Therefore, we get
\bea
B_{zz}&=&B_{\|}\left(\frac{1}{2}
\sin^{2}\theta\sin^{2}\xi+\cos^{2}\theta\cos^{2}\xi\right)+B_{\perp}\left(\frac{1}{2}[
1+\cos^{2}\theta]\sin^{2}\xi+\sin^{2}\theta \cos^{2}\xi\right),\nonumber\\
B_{xx}&=&B_{\|}\left(\frac{1}{2}\sin^{2}\theta[\cos^{2}\phi+\sin^{2}\phi\cos^{2}\xi]
+\cos^{2}\theta \sin^{2}\phi\sin^{2}\xi\right)+B_{\perp}\left(\frac{1}{2}[1+\cos^{2}\theta]
[\cos^{2}\phi+\sin^{2}\phi\cos^{2}\xi]
+\sin^{2}\theta\sin^{2}\phi\sin^{2}\xi\right),\nonumber\\
B_{yy}&=&B_{\|}\left(\frac{1}{2}\sin^{2}\theta[\sin^{2}\phi+\cos^{2}\phi\cos^{2}\xi]
+\cos^{2}\theta \sin^{2}\phi\sin^{2}\xi\right)+B_{\perp}\left(\frac{1}{2}[1+\cos^{2}\theta]
[\sin^{2}\phi+\cos^{2}\phi\cos^{2}\xi]
+\sin^{2}\theta\sin^{2}\phi\sin^{2}\xi\right),\nonumber
\ena
where $\xi$ is the angle between $\bJ$ and $\zhat_{B}$, and $\phi$ is the
azimuthal angle of $\bJ$ in the inertial system $\zhat_{i}$ (see Figure \ref{fig:RFs}). The Larmor precession averaging over $\phi$ gives $\langle\cos^{2}\phi\rangle=\langle\sin^{2}\phi\rangle=1/2$, so that the diffusion coefficients return to those obtained in \cite{1997MNRAS.288..609L}.

\section{Thermal dust emission}\label{apdx:TE}
The intensity of thermal dust emission from dust grains in an optically thin diffuse region can be approximated by
\bea
I_{\rm td}\sim B_{\nu}(T_{d})\tau_{\nu}=B_{\nu}(T_{d})\kappa(\nu)N_{\H},
\ena
where $T_{d}$ is the grain equilibrium temperature that scales with the radiation intensity as $U^{1/5.67}$, and 
\bea
\kappa (\nu) =0.92\times 10^{-25}\left(\frac{\lambda}{250~\mu m}\right)^{-1.8}\sim  10^{-27}\left(\frac{\nu}{100 \rm GHz}\right)^{1.8}\cm^{2} \H^{-1} 
\ena
is the dust opacity per H at long wavelengths \citep{2011A&A...536A..21P}.  Therefore, one obtains 
\bea
\frac{I_{\rm td}}{N_{\H}}\simeq 6.1\times 10^{-18} \left(\frac{\nu}{100 \rm GHz}\right)^{3.8}\left(\frac{T_{d}}{20\K}\right) \Jy\sr^{-1}\cm^{2}\H^{-1}.\label{eq:jtd}
\ena

The polarization of thermal dust emission is observed by \cite{2015A&A...576A.104P}, which has the maximum polarization fraction of 20 $\%$ for the diffuse cloud and decreases with increasing $N_{\H}$. For thermal dust emissivity shown in the present paper, we adopt $T_{d}=20$ K for the CNM, and $T_{d}=60$ K for the RN and PDR. The polarization fraction is fixed to be $20\%$.

\bibliography{ms.bbl}
\end{document}